\documentclass[9pt,twocolumn,twoside]{osajnl}

\journal{jocn} 

\setboolean{shortarticle}{false}
\usepackage{subcaption}
\usepackage{graphicx}
\usepackage{graphics}
\usepackage{amsmath,amssymb,amsfonts}
\usepackage{algorithm}
\usepackage{algpseudocode}

\usepackage{multirow}

\title{RMCSA Algorithm for Congestion-Aware and Service Latency Aware Dynamic Service Provisioning in Software-Defined SDM-EONs}

\author[1*]{Baljinder Singh Heera}
\author[2]{Shrinivas Petale}
\author[1]{Yatindra Nath Singh}
\author[2]{Suresh Subramaniam}

\affil{Department of Electrical Engineering, Indian Institute of Technology Kanpur, Kanpur, Uttar Pradesh, India}
\affil[2]{The George Washington University, Washington, DC, USA}

\affil[*]{baljinder.iitk@gmail.com}

\begin{abstract}
The implementation of 5G and the future deployment of 6G necessitate the utilization of optical networks that possess substantial capacity and exhibit minimal latency. The dynamic arrival and departure of connection requests in optical networks result in particular central links experiencing more traffic and congestion than non-central links. The occurrence of congested links leads to service blocking despite the availability of resources within the network, restricting the efficient utilization of network resources. The available algorithms in the literature that aim to balance load among network links offer a trade-off between blocking performance and algorithmic complexity, thus increasing service provisioning time. This work proposes a dynamic routing-based congestion-aware routing, modulation, core, and spectrum assignment (RMCSA) algorithm for space division multiplexing elastic optical networks (SDM-EONs). The algorithm finds alternative candidate paths based on real-time link occupancy metrics to minimize blocking due to link congestion under dynamic traffic scenarios. As a result, the algorithm reduces the formation of congestion hotspots in the network owing to link-betweenness centrality. We have performed extensive simulations using two realistic network topologies to compare the performance of the proposed algorithm with relevant RMCSA algorithms available in the literature. The simulation results verify the superior performance of our proposed algorithm compared to the benchmark Yen's K-shortest paths and K-Disjoint shortest paths RMCSA algorithms in connection blocking ratio and spectrum utilization efficiency. To expedite the route-finding process, we present a novel caching strategy that allows the proposed algorithm to demonstrate a much-reduced service delay time compared to the recently developed adaptive link weight-based load-balancing RMCSA algorithm. 
\end{abstract}

\setboolean{displaycopyright}{true}

\begin{document}

\maketitle

\section{Introduction}

% Compare latency and blocking performance with cache invalidation frequency = 1, 10, 100, 1000, 10000, with double axis diagram. 

The rapid proliferation of data-intensive applications and the ever-growing demand for high-speed, reliable, and low-latency communication services have thrust optical networking into the spotlight of modern telecommunications \cite{cisco,latency}. Space Division Multiplexing (SDM) has emerged as a transformative technology to meet these escalating demands, offering the potential for immense capacity enhancement in optical fiber communication systems \cite{future,planning}. Elastic Optical Networks (EONs), characterized by their flexibility in accommodating variable bandwidth demands, have gained prominence as a promising platform for realizing the full potential of SDM \cite{flexible,ITU}.

Efficient and intelligent network management is paramount to exploiting the capabilities of EONs in the context of SDM. One of the central challenges faced by operators of such networks is the optimization of routing and spectrum allocation. This problem in SDM-EONs transforms into routing, modulation, core, and spectrum assignment (RMCSA) \cite{RMCSA}. Traditional techniques for solving the RMCSA problem often fall short of fully capitalizing the total capacity offered by the SDM-EON network. The joint-RMCSA problem is rather complex and is broken into sub-problems (routing and modulation assignment problem, core assignment policy, and spectrum assignment problem) to reduce the search space \cite{survey}. For core selection, the first-fit core allocation policy is widely employed because of its simplicity. The spectrum allocation sub-problem suffers due to stringent continuity and contiguity constraints, leading to significant resource fragmentation in networks and service blockage even when resources are available \cite{fragmentation,fairness}. 

In this paper, we will focus on the routing sub-problem, specifically the ability of the routing algorithm to reduce blocking due to non-uniform traffic distribution among networks' links. The performance of the routing and modulation assignment depends on its ability to (i) evenly distribute the network's traffic among the network links, (ii) select the routes over which the highest possible modulation levels can be chosen to enhance the spectrum utilization efficiency, and (iii) choose alternative paths which are also resilient to link-failures.

Traditional benchmark techniques for solving the RMCSA problem often fall short of fully capitalizing the total capacity offered by the SDM-EON network. On the other hand, the state-of-the-art algorithms in the literature that aim to balance load among network links offer a trade-off between blocking performance and algorithmic complexity, which increases service provisioning time. As a result, these highly complex algorithms may not be suitable for future 6G communication services with stringent latency requirements. This chapter presents a dynamic resource allocation algorithm titled "Link-Congestion Aware and Service-Latency-Aware Routing, Modulation, Core, and Spectrum Assignment" (CALA-RMCSA) to address the above limitations. CALA-RMCSA represents a novel approach that blends link congestion and latency awareness into the core of the route-finding process in RMCSA. This algorithm strives to enhance network efficiency, ensuring that the network traffic is distributed judiciously while minimizing latency, which is a critical metric for many contemporary applications, including real-time communication, cloud services, and emerging technologies like the Internet of Things (IoT) and 5G connectivity.

In this work, we embark on a comprehensive journey to introduce and evaluate the proposed CALA-RMCSA algorithm. The remainder of this paper is organized as follows: Section 2 describes the physical-layer model of SDM-EONs and the connection request model for solving the RMCSA problem. In section 3, with an illustrative example, we will shed light on link-congestion problems due to link-betweenness centrality in the RMCSA algorithm. Section 4 talks about the earlier studies that have been done in the literature to address the link-congestion issue in SDM-EONs. Thereafter, we describe the motivation behind this work and novel contributions. Subsequently, In section 5, we detail the principles and components of the CALA-RMCSA algorithm, elucidating how it integrates link congestion and latency awareness. We then present an extensive evaluation of the algorithm's performance using simulation experiments in section 6, demonstrating its superior capabilities compared to the existing benchmark and state-of-the-art algorithms. Finally, section 7 concludes the paper by summarizing the key contributions and discussing the avenues for future research.

\section{System Model}

\subsection{Physical Layer Model}
A graph G(V, E, C, S) can be used to illustrate an optical network. The set of nodes, in this case, is represented by V, the set of links by E, the set of cores by C, and the set of spectrum slices that are available in each core by S. A single spectrum slice has a bandwidth of $B_s$. A link $l\in E$ has a link length $L_l$ in km, and each link has the same bandwidth $B = B_s\times |S|\times |C|$. The nodal degree of a node $n\in V$ is indicated by $D_n$. In the network, the set of all node pairs is represented by $Z$, where $|Z| = |V|(|V|-1)$. Average link length ($L_{avg}$), average nodal degree ($D_{avg}$), and link-betweenness centrality ($LBC$) are a few important physical characteristics of the network's topology.

\subsection{Request Model and RMCSA Problem}

In a network, a lightpath request is denoted by $r(s,d,b)$. For the holding period $T_h$, a new request requires the establishment of a dedicated connection with a desired data rate of $b$ Gbps between source node $s$ and destination node $d$. The set $\delta$ consists of a collection of feasible data rates, and the EON can establish the lightpath at any of the pre-defined data rates from $\delta$. EON uses distance-adaptive modulation formats to improve spectrum efficiency. The spectral efficiency in bits/sec/Hz can be increased by transmitting more bits per symbol using the highest-possible modulation formats.  In SDM-EONs, the RMCSA algorithm is employed to allocate resources to arriving requests. When a new request arrives, the algorithm finds a physical path between $s$ and $d$ nodes, which may span multiple links. Depending on the length of the path ($L_p$) and $b$, the RMCSA algorithm computes the minimum number of spectrum slots which are required to support the requested data rate as,
 
 \begin{equation} \label{eqn_S}
  SS_r = \left\lceil\frac{b}{2\times B_s \times m}\right\rceil.
\end{equation}

Here, $m$ indicates the spectrum efficiency of the modulation format utilized in bits/symbol. Dual-polarization multiplexing reduces $SS_r$ by another factor of two by doubling the sub-channel's bandwidth. After calculating $SS_r$, the connection request can also be expressed as $r(s,d,SS_r)$. The RMCSA algorithm searches for a route with the necessary slots available. A lightpath is established between the $s$-$d$ pair after identifying the resources. The arrival and departure of connection requests occur dynamically in a real traffic scenario. In accordance with $T_h$, the occupied resources are freed and made available for upcoming requests. The goal of the RMCSA algorithm is to minimize service interruptions caused by unavailable resources while adhering to spectrum continuity, spectrum contiguity and core continuity constraints.  

\section{Link-betweenness Centrality}\label{section2}

 The fraction of shortest routes passing via a link relative to the total number of shortest paths between all source-destination pairs in a network is known as the link betweenness centrality ($LBC$) of that link \cite{LBC}. In an undirected graph, the total number of $s$-$d$ pairs equals $|V|(|V|-1)/2$, where V is the total number of nodes in the network. $LBC$ of a link $l$ in a network is defined as $$LBC_l = \frac{\sigma_l}{\sigma}$$
 Here, $\sigma$ denotes the total number of shortest paths while $\sigma_l$ represents the number of shortest paths that traverse through link $l$. Using $LBC$, we can determine which links are less central and which are more central. Thus, we can infer more probable bottleneck links.

\begin{figure}[htbp]
\centering
\includegraphics[width=0.99\linewidth]{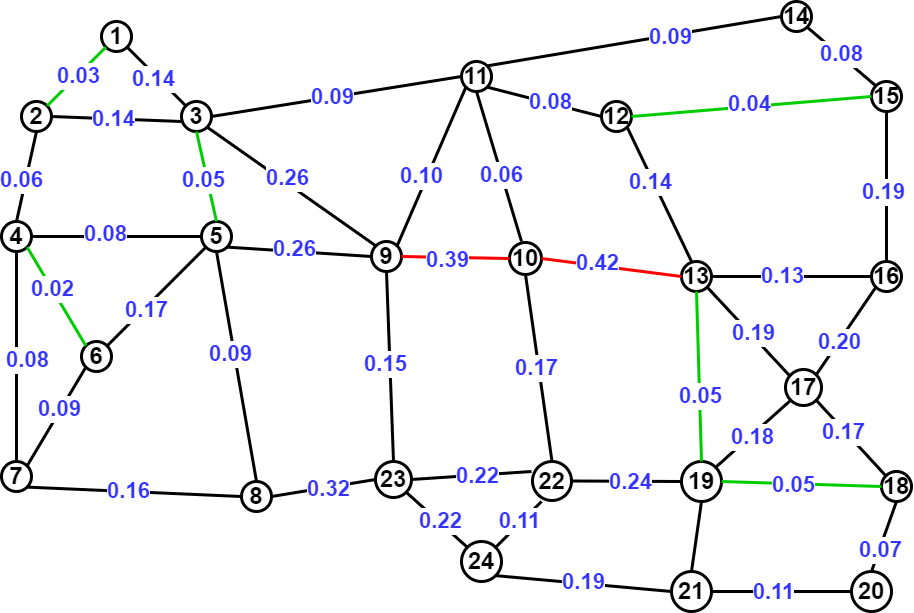}
\caption{Link betweenness centrality of USNET networks' links.}
\label{fig:centrality}
% \hspace{1mm}
% \rule{8.8cm}{0.1mm}
\end{figure} 

Fig. \ref{fig:centrality} provides the $LBC$ of each link in the USNET \cite{ondm} topology. Out of all the links in the networks, links $\{9, 10\}$ and $\{10, 13\}$ (marked as red) have the highest centrality, whilst a few other links (marked as green) have the lowest centrality. Due to spectrum continuity limits, if we use a single shortest-path routing method in the RMCSA algorithm, the central links may fill up quickly and begin to block multi-hop connection requests whose shortest paths pass through them. Other non-central links, on the other hand, might continue to be underutilized since the shortest routes will not frequently select them. The larger the standard deviation value in $LBC$ of all links ($\sigma_{LBC}$), the more challenging it will be for the network operators to fully utilize the resources at their disposal. Lower network resource utilization at their end could raise the price of the services they provide.     

\section{Relevant Works and Paper Contributions}

This section provides an overview of the current research reported in the literature, which aims to reduce the blocking due to $LBC$, thus improving spectrum utilization efficiency by balancing the traffic load among network links.

To solve the routing-sub problem, the shortest path algorithm is widely employed in routing and spectrum allocation algorithms. The motivation behind using shortest-length paths is their ability to utilize higher-order modulation formats, thus increasing overall spectrum efficiency in bits/symbol \cite{jinno2010distance}. Moreover, the pre-computed shortest paths are the easiest and fastest method for route discovery. Dijkstra's shortest path (SP) algorithm \cite{szczesniak2019generic} can be implemented to find and store in memory the shortest routes between each $s$-$d$ pair in a network. By avoiding the requirement to calculate the route for every request, the \textbf{SP-RMCSA} algorithm \cite{fairness} operates significantly faster. However, shortest path routing algorithms fall short of evenly distributing the networks' traffic among all network links. As described in section \ref{section2}, when there is a significant disparity in the $LBC$ of the comprising links, employing a single shortest path leads to over-utilization of some links, which in turn increases the blocking of multi-hop connection requests. Since most realistic network topologies have links with varying link lengths and are non-uniform in link and nodal degrees, the shortest path algorithm policy causes network congestion because of bottlenecks caused by a few most central links \cite{ondm}. Furthermore, the SP-RMCSA looks for resources only on the shortest paths, whereas other alternative paths may be present in the networks between the same $s$-$d$ pair with available resources.

Then, Yen's K-shortest paths-based \textbf{KSP-RMCSA} algorithm comes into the picture \cite{Yen,RMCSA} that iteratively determines if resources are available on any of the first K candidate shortest paths. Our prior work \cite{ondm} examined and demonstrated how the KSP-RMCSA algorithm performs better with increasing K despite the fact that its performance reaches a saturation point for larger values of K.  As a result, the KSP-RMCSA algorithm performs better in connection blocking and resource utilization efficiency with higher computational complexity. Since the algorithm checks for the availability of required spectral resources until all K-alternative paths are exhausted, it takes longer to set up lightpaths. However, the K-alternative shortest paths between each $s$-$d$ pair in a network can be pre-computed and stored to speed up the route-finding process. Just like the shortest-path algorithm, the KSP-RMCSA algorithm concentrates on finding the routes with the least amount of bandwidth usage, ignoring the traffic load distribution. Therefore, the KSP algorithm might also result in bottlenecks in scenarios with dynamic traffic choosing similar links more frequently. Since the alternative paths are not required to be disjoint in Yen's KSP algorithm, the K-alternative paths identified may share some links. The connection request will be denied on both of the alternate paths provided by the KSP algorithm if they share central or congested links.

A routing algorithm based on K-disjoint pathways (KDP) has been employed in certain works to address the problem of common links in alternative paths in the RMCSA problem \cite{teresa}. Disjoint paths-based routing is also highly popular in the designing of resilient optical networks \cite{resilient,disaster}. As the name implies, each candidate path must be link-disjoint because it provides resilience in the event of a link failure. There won't be a common link among K-alternative paths. However, the candidate paths found by the KDP algorithm are often longer than the alternative shortest paths found by Yen's KSP algorithm. This is because the KSP algorithm finds the shortest paths by first setting the weights of each link in the shortest path to infinity one at a time to identify an alternate path. Next, the alternate routes are listed according to increasing path length. However, the Disjoint-path routing technique sets the cost of all the links in the shortest path to infinity and determines the next shortest path to ensure that the alternative path is link-disjoint to the shortest path \cite{disjoint}. It may, therefore, select longer routes which support lower order modulation formats, necessitating high $SS_r$ to set up a lightpath with bandwidth demand $b$. In addition, the disjoint path identification algorithm removes all of the links in the shortest paths without contemplating which link created obstruction. However, some of the links in the shortest pathways might have adequate slots available. The fact that nodal degrees restrict the availability of disjoint paths between $s$-$d$ pairings in a network is another drawback of employing a disjoint-path routing algorithm. The maximum possible number of disjoint routes between $s$ and $d$ is limited to $min$($D_s$, $D_d$), where $D_s$ and $D_d$ represent the nodal degrees of the source and destination nodes, respectively. Fig. \ref{topologies} displays two real network topologies, demonstrating many nodes with nodal degree two. There are only two disjoint paths that may be found if these nodes are either source or destination. As a result, we discovered that the \textbf{KDP-RMCSA} algorithm, which is constrained by nodal degrees of network links, attempts to circumvent the issue of link congestion at the expense of lengthier alternative paths with scarce availability.

The aforementioned routing strategies attempt to minimize connection blocking by checking the availability of required resources on K alternative paths, albeit at the expense of greater computational complexity and/or less efficient modulation formats. None of these techniques use information on the spectrum occupancy status of network links to identify alternative routes with a higher acceptance probability. Finding the least loaded route from the source to the destination is a driving principle for determining the load-balanced paths in real time. Thanks to developments in software-defined networking (SDN), the central controller can now access real-time network data, including traffic load \cite{SDN,SDN2}. As a connection request comes in, the route-finding algorithm can take advantage of the real-time traffic data and can identify the least loaded path to minimize the request blocking. 

A load-balancing routing and modulation assignment (LB-RMA) based RMCSA algorithm was proposed by Zhang \textit{et al.} in \cite{LBFA}, utilizing the traffic information available at the SDN controller. Their routing method is also greedy, but instead of using connection lengths as link weights, it uses the traffic load on each link ($TL_l$) as its link weight. 
$$TL_l = \sum_{\forall c} S_o^c$$
$S^c_o$ denotes the number of slots occupied in core $c$ on the link $l$. Since a link's number of occupied spectrum slots fluctuates, so does its weight. By successfully lowering the service blocking ratio, their suggested algorithm improves the efficiency of network utilization. However, the load-balancing routing method may become unstable with high traffic loads since it must update weights too frequently. The average increase in service delay caused by the periodic weight update mechanism, which is crucial to latency-aware future 5G and 6G communication standards, was not studied in their work. 

An improved version of the load-balancing RMCSA algorithm was proposed by Lan \textit{et al.} \cite{weights}, wherein the new link weight ($NLW_l$) is jointly determined by normalized link length and traffic load on the link in terms of spectrum occupancy ratio.
$$NLW_l = \alpha*\left(\frac{L_l}{L_{max}}\right) + (1-\alpha)*\left(\frac{\sum_{\forall c} S_o^c}{|S|\times |C|}\right)$$
$L_l$ and $L_{max}$ denote the length of link $l$ and the length of the longest link in the network, respectively. Their improved load-balancing \textbf{LB-RMCSA} algorithm leads to further improvements in blocking performance and resource utilization efficiency. They did not, however, present the relationship between the ideal weight coefficients ($\alpha$) and network characteristics; instead, they found the weight coefficients' optimal value through repeated experiments for every network topology. However, since the link's occupancy status changes after providing resources to each connection request and after releasing resources after the service's holding time $T_h$ passes, their technique still requires periodic updating of the link weights.

The requirement for periodic updating of the link weights in load-balancing RMCSA algorithms presents challenges in designing a faster route-finding algorithm because (i) we cannot pre-compute the routes between each $s$-$d$ node pair and store them in memory, and (ii) if the algorithm computes the routes in real-time, the previously computed shortest routes between a specific $s$-$d$ node pair may not be the shortest routes between the same $s$-$d$ pair after updating of the link's weights in future. The motivation behind our current work is to develop a cognitive heuristic algorithm that uses traffic data to identify routes with high acceptance probability while avoiding the need for a periodic weight update, thus enabling the discovery of alternative paths more quickly than the LB-RMCSA algorithm to minimise service latency.

Chatterjee \textit{et al.} \cite{batch} have presented a batch-processing-enabled proactive-fragmentation-aware strategy for fair resource distribution to diverse requests in SDM-EONs. Their proposed algorithm achieved near-perfect fairness by prioritizing the requests based on the number of hops the lightpath spans and the number of spectrum slots it requires. However, in contrast to immediate request fulfillment, the batch-forming process may result in appreciable processing delays, especially at lower traffic loads. Since computation is becoming cheaper, research is also focusing on developing machine learning (ML) assisted resource allocation techniques to optimize the RMCSA problem. To optimize the trade-off between spectrum utilization and crosstalk tolerance in MCF-enabled SDM-EONs, Petale \textit{et al.} \cite{ML} have proposed an ML-aided optimization strategy that significantly reduces the spectrum wastage when compared to other RMCSA algorithms. The additional delay introduced by the highly complex ML-aided RMCSA algorithms further needs to be investigated.

In this paper, we present a dynamic routing-based resource allocation algorithm for SDM-EONs that is cognizant of network congestion, service delay, and resilience. The preliminary results of our proposed algorithm were presented in a conference \cite{ondm}, where we showed that our congestion-aware routing algorithm with only two alternative paths outperforms the benchmark KSP-RMCSA algorithm having four alternative paths in terms of both request blocking probability and resource utilization efficiency. However, the proposed CA-RMCSA algorithm in \cite {ondm} could only compute a maximum of two candidate routes. Furthermore, the proposed algorithm is slower in discovering alternative routes than the KSP RMCSA algorithm because it does not incorporate latency awareness and requires run-time computations to compute congestion-aware alternative paths. 

We have performed extensive simulation experiments and performance comparisons with shortest-path routing based \emph{SP-RMCSA} algorithm, K-shortest paths routing based \emph{KSP-RMCSA} algorithm, K-Disjoint paths routing based on \emph{KDP-RMCSA} algorithm, and adaptive link weights-based load-balancing \emph{LB-RMCSA} algorithms. We will also demonstrate a novel caching method to save pre-computed paths in memory and reuse them to minimize service latency. Our proposed routing algorithm also exhibits resilience in case of single link failures. Along with request blocking performance and resource utilization efficiency, we introduce a new metric to compute the average service delay offered by each RMCSA algorithm under consideration. This metric would also be helpful for the network operators to assess the performance of each algorithm in terms of service latency.

The novel contributions added to the present work are the following.

\begin{itemize}
    \item Improved congestion-aware routing algorithm is not restricted to finding only two alternative candidate paths \cite{ondm}. It can find any number of available congestion-aware alternative candidate paths between a $s$-$d$ pair.
    \item The proposed dynamic routing algorithm also ensures resiliency in case of single link failure.
    \item  We propose and demonstrate a novel caching method to save pre-computed paths in memory and reuse them to find paths for future connection requests. The implemented cache helps to hasten the dynamic route-finding process significantly.
    \item Along with service blocking performance and resource utilization efficiency, we introduce two additional metrics to compute the average service delay offered by each RMCSA algorithm under consideration. These metrics would be helpful for the network operators to also assess the performance of each algorithm in terms of service latency.
    \item In previous work, we compared the performance of our proposed algorithm with the benchmark KSP-RMCSA algorithm. However, in this work, we present a performance comparison with other benchmark and state-of-the-art routing algorithms such as SP-RMCSA, KDP-RMCSA, and LB-RMCSA algorithms along with the KSP-RMCSA algorithm.
    \item We test the performance of the underlined algorithm on two different realistic networks having different physical attributes ($D_{avg}$ and $\sigma_{LBC}$) and try to find the relation in performance of aforementioned algorithms with networks' physical attributes.
\end{itemize}

% \section{Corresponding author}

% We require manuscripts to identify a single corresponding author. The corresponding author typically is the person who submits the manuscript and handles correspondence throughout the peer review and publication process. If other statements about author contribution and contact are needed, they can be added in addition to the corresponding author designation.

%Example with the corresponding author designated by an asterisk:

%\author{Author One\authormark{1} and Author Two\authormark{2,*}}

%\address{\authormark{1}Peer Review, Publications Department,
%Optica Publishing Group, 2010 Massachusetts Avenue NW,
%Washington, DC 20036, USA\\
%\authormark{2}Publications Department, Optica Publishing Group,
%2010 Massachusetts Avenue NW, Washington, DC 20036, USA\\
%%\authormark{3}xyz@optica.org}

%\email{\authormark{*}opex@optica.org}}

%Example with the corresponding author designated by an asterisk and a note indicating equal contributions by two authors.

%\author{Author One\authormark{1,3} and Author %Two\authormark{2,3,*}}

%\address{\authormark{1}Peer Review, Publications Department,
%Optica Publishing Group, 2010 Massachusetts Avenue NW, %Washington, DC 20036, USA\\
%\authormark{2}Publications Department, Optica Publishing Group, %2010 Massachusetts Avenue NW, Washington, DC 20036, USA\\
%\authormark{3}The authors contributed equally to this work.\\
%\authormark{*}opex@optica.org}}

\begin{table}[htbp]
\centering
\caption{\bf List of Notations and Symbols}
\renewcommand{\arraystretch}{1.25} % Default value: 1
\begin{tabular}{ll}
\hline
Notation & Description \\
 \hline
% & \bf Given Parameters\\
G(V, E, C, S) & Representing network topology\\ 
$V$ & Set of nodes in the network\\
$E$ & Set of links in the network\\
$C$ & Set of cores in the MCFs\\
$S$ & Set of spectrum slots in a core\\
$b$ & Bandwidth requested by a CR\\
$SS_r$ & Spectrum slots required by a CR\\
$m$ & Modulation format\\
$\alpha$ & Data rate supported by a modulation format \\
$D_{avg}$ & Average nodal degree of a network\\
$SOR_l$ & Spectrum Occupancy Ration of link $l$\\
$l_{max}(P_k)$  & link in path $P_k$ with highest SOR \\
$L_{avg}$& Average Link length in a network\\
$LBC_l$ & Link Betweenness Centrality of a link $l\in E$\\
$\sigma_{LBC}$ & Standard Deviation of LBC of all links \\
$R_b$ & Total requests blocked \\
$R_a$ & Total requests accepted \\
$T_h$ & Holding time of a connection request \\
$H_p$ & Number of hops in a working path $P$ \\
$\tau$ & Total observation time \\
$L_p$ & Path length of working lightpath\\
$L_e$ & List of edges to be excluded \\
\hline
\end{tabular}
  \label{tab:symbols}
\end{table}

\section{Proposed Algorithms}
\label{section5}

\begin{algorithm}[hb]
    \caption{To find the alternative candidate routes.
    }
    \begin{algorithmic}[1]
        \Procedure{Cong. Aware-Alternative Path (CA-AP)}{}\newline
        \textbf{Input:} $G(V,E,C,S)$, $s$, $d$, $P_{k-1}$, $l_{max}(P_{k-2})$,.....$l_{max}(P_{k-(k-1)})$ \newline
        \textbf{Output:} $P_{k-CA}$
        \State Fetch the links of the $P_{k-1}$
        \State For every link $l \in P_{k-1}$, compute $SOR_l$ using Eqn. \ref{eqn:SOR}.
        \State Fetch the link with $max(SOR_{k-1})$ and set it $l_{max}(P_{k-1})$
        \State Set the weight of $l_{max,k-1}$, $l_{max,k-2}$,......$l_{max,k-(k-1)}$ = $\infty$
        \State Find shortest path ($P_{k-CA}$) using \textit{Dijkstra's algorithm} 
        \State Set $l_{max,k-1}$,... $l_{max,k-(k-1)}$ equal to original weights
        \State \textbf{Return}: $P_{k-CA}$.
        \EndProcedure 
    \end{algorithmic}
    \label{alg:CA}
\end{algorithm}

\subsection{Congestion-Aware Alternative Paths}

We now present the congestion-aware alternative path (CA-AP) routing strategy used in the proposed CALA-RMCSA algorithm, which is illustrated in Algorithm \ref{alg:CA}. If a request is denied on the first shortest path—that is, if the RMCSA algorithm is unable to discover the required slots in any core along the shortest path between $s$ and $d$—the CA-AP algorithm is triggered. The algorithm calculates the spectrum occupancy ratio ($SOR_l$) for each link in the shortest path $P_{1s}$ as, 

\begin{equation} \label{eqn:SOR}
  SOR_l = \frac{\sum_{\forall c} S_o^c}{|S|\times |C|}
\end{equation}

The algorithm then determines the link, $l_{max}$, which has the highest value of $SOR$. Then, using Dijkstra's algorithm, the algorithm determines the shortest path between $s$ and $d$ node after temporarily setting the weight of the $l_{max}$ link equal to $\infty$. The algorithm returns the path found as a $2^{nd}$ congestion-aware alternative path $P_{2-CA}$ and resets the weight of the $l_{max}$ link to its actual value. Similarly, if a request is blocked on all (K-1) candidate paths, the algorithm finds the $K^{th}$ congestion-aware alternative path (lines 5 and 6 in Algorithm \ref{alg:CA}) after temporarily removing each highest congested link in all (k-1) paths.   

\begin{figure}[htbp]
\centering
\includegraphics[width=0.99\linewidth]{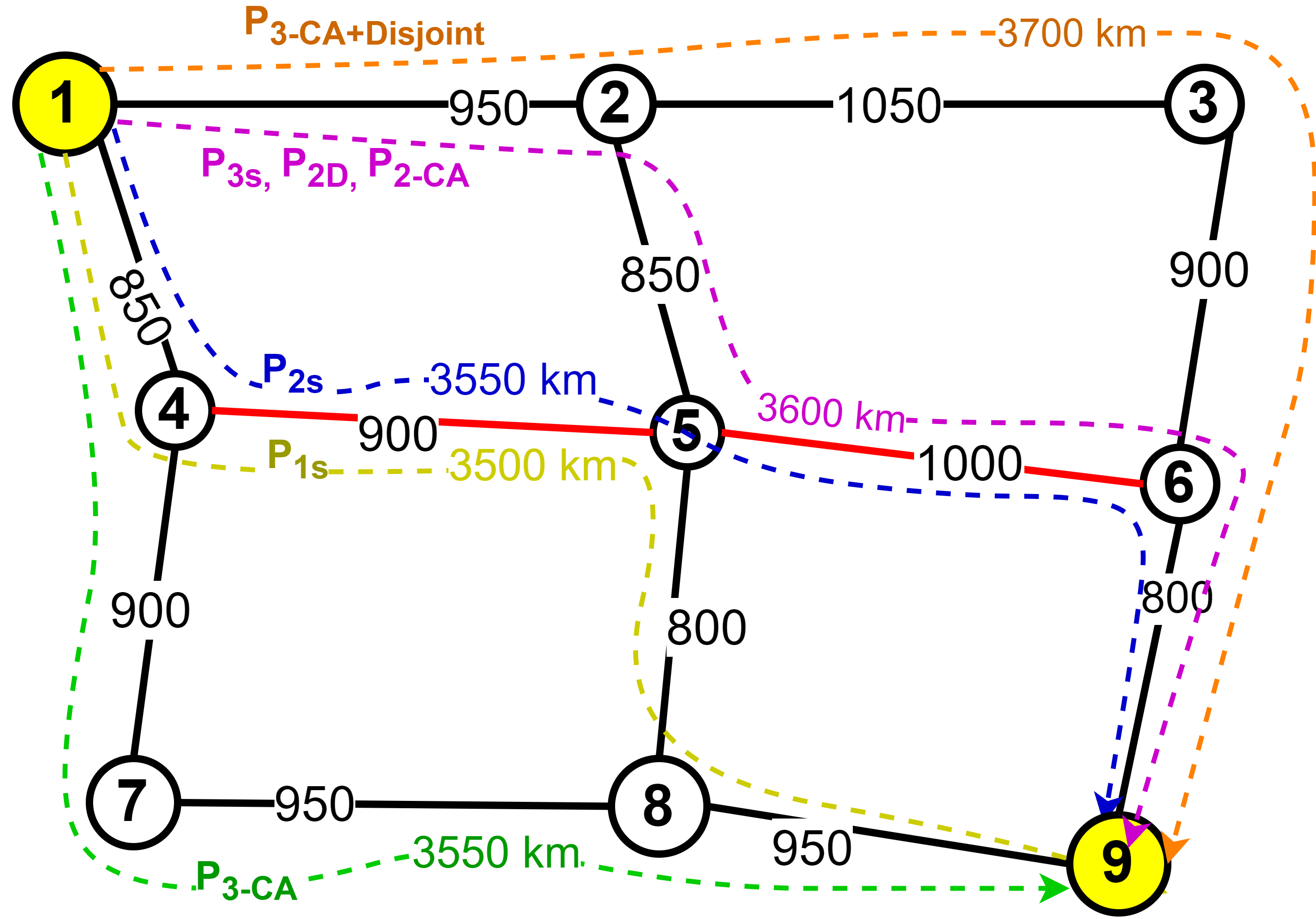}
\caption{Illustrative example of route finding of KSP, KDP and CA-AP algorithms.}
\label{fig:example2}
% \hspace{1mm}
% \rule{8.8cm}{0.1mm}
\end{figure}

Utilizing Fig. \ref{fig:example2} as an example, we elucidate the $CA-AP$ algorithm and compare it with KSP and KDP routing algorithms. There are twelve links with link lengths in kilometres and nine nodes in the test network shown in Fig. \ref{fig:example2}. We assume that some non-central links in the networks have plenty of available slots, while the most central links, links $\{4,5\}$ and $\{5,6\}$, are highly congested at the moment. Let's say a new request comes in that asks to establish a lightpath from node $1$ to node $9$. Assume for the moment that K=3.

The KSP algorithm will check the availability of required resources on first three shortest paths. It is apparent that link $\{4,5\}$, which is congested, is shared by both the first shortest path ($P_{1s}$) and the second shortest path ($P_{2s}$). If a request gets blocked on the path $P_{1s}$ (marked yellow), it will get blocked on  $P_{2s}$ (marked as blue) too, which also spans the same congested link $\{4,5\}$. Subsequently, the algorithm will check resource availability on the third shortest path $P_{3s}$ (marked as pink). The request will again get block on  $P_{3s}$ since it shares the other congested link $\{5,6\}$ with $P_{2s}$. 

On the other hand, in the event that the request is blocked on $P_{1s}$, the KDP routing algorithm will find the next shortest path, $P_{2D}$ (shown as pink), which is the link-disjoint to $P_{1s}$. The request will be blocked on $P_{2D}$, because both $P_{1s}$ and $P_{2D}$ contain congested links $\{4,5\}$ and $\{5,6\}$, respectively. Next, the third link-disjoint path ($P_{3D}$) between nodes 1 and 9 will be searched by the KDP algorithm. Nevertheless, a third link-disjoint shortest path $P_{3D}$ does not exist because both nodes have a nodal degree $D<3$. As a result, the request will get blocked.  

The proposed CA-AP algorithm determines that link $\{4,5\}$ has the highest $SOR$ after the request is blocked on $P_{1s}$. The second alternative path $P_{2-CA}$ is found by removing link $\{4,5\}$, as explained in Algorithm \ref{alg:CA}. This ensures that the $P_{2-CA}$ does not contain the bottleneck link. There is another congested link $\{5,6\}$ in $P_{2-CA}$, which will lead to resource unavailability. Then, in order to determine the third alternative path, the algorithm will eliminate links $\{4,5\}$ and $\{5,6\}$, which are the busiest links in paths $P_{1s}$ and $P_{2-CA}$, respectively. The third alternative path ($P_{3-CA}$) leads to success as the route does not include any of the bottleneck links. Eventually, the proposed CA-AP routing algorithm finds a route bypassing the bottleneck links without necessitating link-disjoint property, as required by the KDP algorithm, or periodic updates of link weights against $SOR$, as required by the LB-RMCSA algorithm. 

\subsection{Congestion-Aware $K^{th}$ Disjoint Path}

\begin{algorithm}[htbp]
    \caption{To find $K^{th}$ CA and disjoint candidate path.
    }
    \begin{algorithmic}[1]
        \Procedure{Congestion-Aware Disjoint Path (CA-D)}{}\newline
        \textbf{Input:} $G(V,E,C,S)$, $s$, $d$, $P_1$, $l_{max}(P_2)$,.....$l_{max}(P_{k-1})$ \newline
        \textbf{Output:} $P_{K-CA-D}$
        \State Fetch the links of the $P_1$ between $s$-$d$ pair
        \State Set weight of all links in $P_1$, $l_{max}(P_2)$,....,$l_{max}(P_{K-1})$ = $\infty$
        \State Find shortest path ($P_{K-CA-D}$) using \textit{Dijkstra's algorithm} 
        \State Revert weights of links in $P_1$, $l_{max}(P_2)$,....,$l_{max}(P_{K-1})$ to original weights
        \State \textbf{Return}: $P_{K-CA-D}$.
        \EndProcedure 
    \end{algorithmic}
    \label{alg:Dis}
\end{algorithm}

Algorithm \ref{alg:CA} does not guarantee link-disjointness among alternative paths. However, the preference for link-disjoint alternative paths is evident in designing resilient routing algorithms. To address limitations in the KDP-RMCSA algorithm, primarily imposed by the average nodal degree in the network, network operators may find it necessary to introduce new links. This, however, poses a potential increase in capital expenditure (CapEx). The investigation conducted by the authors on various real network topologies, such as the Europe Network (Figure \ref{topologies} (a)), German Network (Figure \ref{topologies} (b)), USNET \cite{ondm}, NSF-14 \cite{fairness}, Indian-RailTel \cite{fairness}, JPN-12 \cite{RMCSA}, reveals that each node in these topologies possesses a nodal degree $D$ $\geq$ 2. Consequently, finding two link-disjoint alternate paths in real network topologies becomes feasible. Leveraging this observation, we propose a modification to our routing algorithm. After the first $K-1$ alternative paths are exhausted, we use \textbf{Algorithm \ref{alg:Dis}}  to find the $K^{th}$ candidate paths. Algorithm \ref{alg:Dis} ensures that the $K^{th}$ congestion-aware candidate path is link-disjoint from the first shortest path. In the illustrative example shown in Figure \ref{fig:example2}, the $K^{th}$ candidate path ($P_{3-CA-D}$) is a third congestion-aware alternative path which is also link-disjoint to the shortest path $P_{1s}$.

\subsection{Novel Caching Method to Speed-up the Route Finding Process}
\label{caching}
This section describes how we use cache to hasten the real-time route-finding process. We use $memoization$ to speed up the dynamic route-finding process. \emph{Memoization} is a caching technique used in computer programming to expedite the computer program by storing the output results of computationally complex functions in a cache and returning the results when the same input occurs again. We store the pre-computed shortest paths as well as congestion-aware alternative paths in memory using a dictionary data structure whose key comprises source node $s$, destination node $d$, and the links that are to exclude $L_e$. $L_e$ remains null in case of storing the first shortest path between a $s$-$d$ pair, computed using Dijkstra's algorithm. The algorithms store the key as a Tuple ($s,d,L_e$). $L_e$ denotes the set of links which are set equal to $\infty$ before finding a path between the $s$-$d$ pair. The value against a key is the shortest path between $s$-$d$ pair in the network computed by excluding all links in $L_e$.

\begin{figure}[htbp]
\centering
\includegraphics[width=0.8\linewidth]{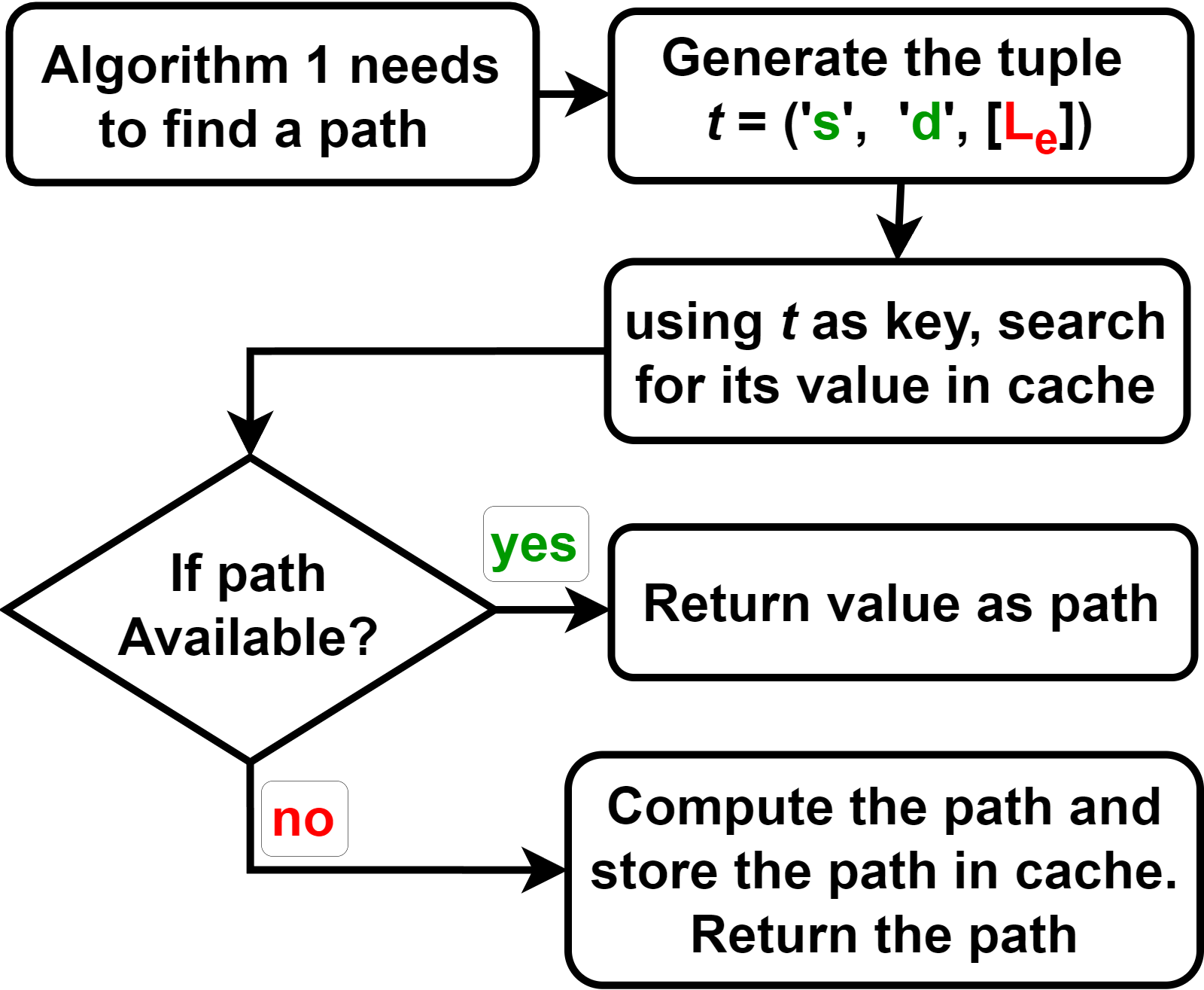}
\caption{Flowchart of the steps involved in implementing cache.}
\label{fig:cache}
% \hspace{1mm}
% \rule{8.8cm}{0.1mm}
\end{figure}

The steps involved in implementing and utilizing the cache are given in the flowchart, illustrated in Fig. \ref{fig:cache}. When the CALA-RMCSA algorithm needs to find the shortest path between any $s$-$d$ pair, or it needs to find the congestion-aware alternative path using the Algorithm \ref{alg:CA}, at first, it checks if the required path is already available in the cache using its key ($s,d,L_e$). If the required path is available, it simply fetches it from the cache. Otherwise, the algorithm computes the required path (using Dijkstra's shortest path algorithm if $L_e$ = null, or using Algorithm \ref{alg:CA} if the resources are unavailable on the shortest path) and stores the same in the cache.

\subsection{Congestion-Aware and latency Aware Resilient CALA-RMCSA Algorithm}

\begin{algorithm}[ht]
    \caption{Main Algorithm}
    \label{alg:main}
    \textbf{Input:} $G(V,E,C,S)$, $r(s,d,b)$, $k$ \\
    \textbf{Output:} $P_k$, c, $SS_r$, SSI, $m$\hfill
    % \Comment{$\%${ $m$ chosen using Table \ref{tab:modulation}}}
    \begin{algorithmic}[1]
        \For {each request $r$}
        % \State Initialize $k = 1$, $c$ = 1
             \For{$k$ = 1 to $K$}   
                \State Fetch the shortest path $P_1$ between $s$-$d$ pair
                \State Select $m$ against $L_p$ using Table \ref{tab:modulation}
                \State Compute $SS_r$ using Eqn. \ref{eqn_S}
                    \For{$c$ = 1 to $|C|$}
                        \State Find $SS_r$ in the $c^{th}$ core
                        \If {$SS_r$ available}
                            \State \textbf{return} $P_k$,c, $SS_r$, SSI
                        % \EndIf
                        \ElsIf{ i $<$ $|C|$}
                            \State $c$ = $c$ + 1, return to step 7
                        \EndIf
                        \State \textbf{end if}
                    \EndFor
                    \State \textbf{end for}
                \State $k$ = $k$ + 1
                \If{$k$ $<$ $K$}
                    find $P_{k-CA}$ using Algorithm \ref{alg:CA}
                    \State return to step 4
                \ElsIf{$k$ $=$ $K$}
                    \State find $P_{K-CA-D}$ using Algorithm \ref{alg:Dis}
                    \State return to step 4
                \EndIf
                \State \textbf{end if}
            \EndFor
            \State \textbf{end for}
        \EndFor
        \State Block the request
        \State \textbf{end for}
    \end{algorithmic}
\end{algorithm}

The proposed CALA-RMCSA algorithm is given in Algorithm \ref{alg:main}. For each new request $r(s,d,b)$ that arrives at the SDN controller, the algorithm finds the shortest path between the source and destination nodes using Dijsktra's algorithm. According to the shortest path length, the highest possible modulation level is chosen using Table \ref{tab:modulation}. The number of required spectrum slots $SS_r$ are computed using Eqn. \ref{eqn_S}. The algorithm then checks the availability of required slots in MCF cores one at a time using a first-fit policy. If the resources are unavailable in any of the cores in the shortest path, the algorithm finds the congestion-aware alternative path using Algorithm \ref{alg:CA}, thus checking if the resources are available in any of the cores in alternative paths until all $K-1$ paths are exhausted. When the first $K-1$ paths are exhausted, the algorithm finds $K^{th}$ alternative paths using the Algorithm \ref{alg:Dis}, which should be link-disjoint with the first shortest path ($P_1$). If the resources are found on a route satisfying the continuity and contiguity constraints, the algorithm returns path $P_k$, core index $c$, starting slot index $SSI$, and required spectrum slots $SS_r$. A lightpath is established between $s$ and $d$ node by occupying $SS_r$ slots starting from $SSI$ in core $c$ of path $P_k$ for holding time $T_h$. If the algorithm cannot find the resources in any of the alternative candidate paths, the request is considered blocked, and the algorithm proceeds to process the next request.    

\section{Simulation Experiments and Results Discussion}

\begin{table}[htbp]
\centering
\caption{\bf Network Parameters}

\renewcommand{\arraystretch}{1.25} % Default value: 1

\begin{tabular}{c|ccc}
\hline
\textbf{Parameters}  & Europe & German \\
\hline
$|V|$  &28 & 17\\
\hline
$|E|$& 41& 26\\
\hline
$D_{avg}$ & 2.93 &3.05\\
\hline
$L_{avg}$ &625.7 & 170.3\\
% \hline
% $\sigma_{LBC}$ & 41.2 & 19.85 & 25.3\\
\hline
\end{tabular}
  \label{tab:nru}
\end{table}

\subsection{Simulation Environment}

\begin{figure} [ht]
\captionsetup[subfigure]{justification=centering}
    \centering
      \begin{subfigure}{0.35\textwidth}
        \includegraphics[width=\textwidth]{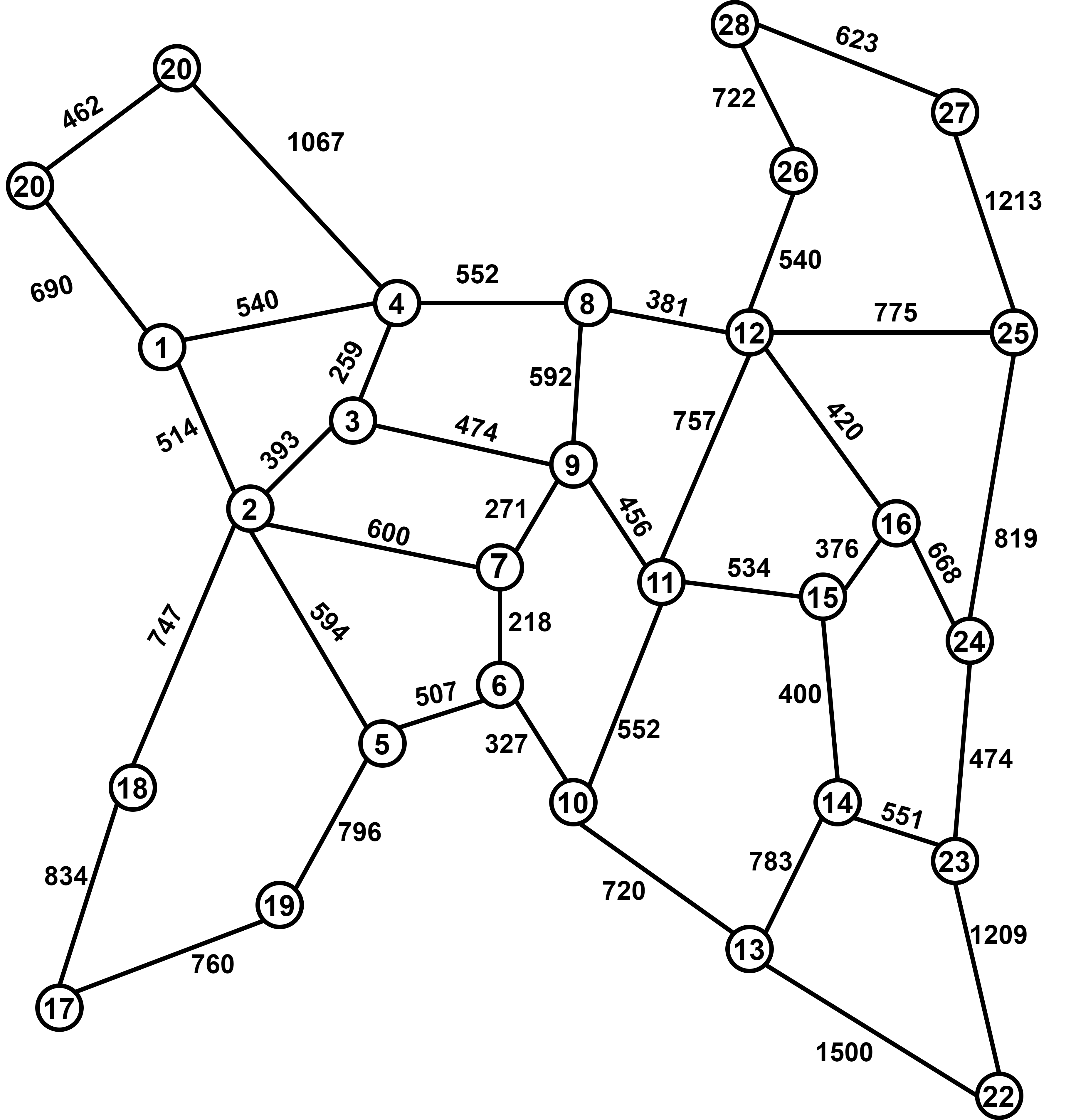}
          \caption{}
          \label{subfig:euro}
      \end{subfigure}
      \begin{subfigure}{0.30\textwidth}
        \includegraphics[width=\textwidth]{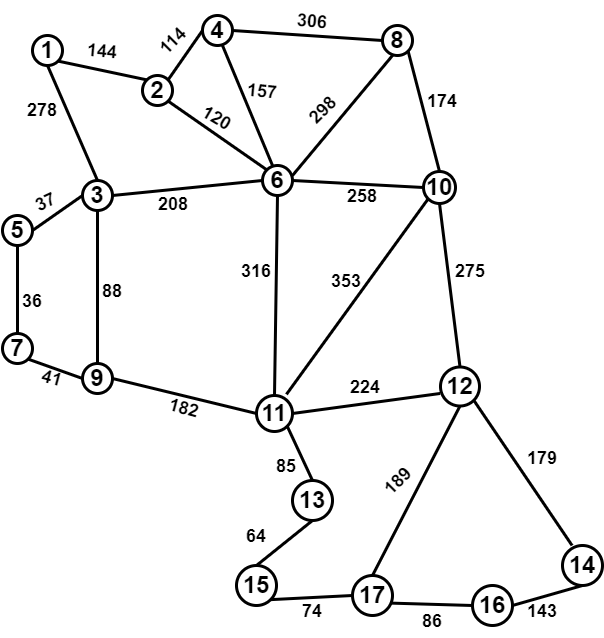}
          \caption{}
          \label{subfig:german}
      \end{subfigure}
      % \begin{subfigure}{0.38\textwidth}
      %   \includegraphics[width=\textwidth]{USNET_New.png}
      %     \caption{}
      %     \label{subfig:usnet}
      % \end{subfigure}      
    \caption{Network topologies with link length in km: (a) Europe Network, and (b) German Network}
\label{topologies}
% \rule{8.3cm}{0.1mm}
\end{figure}

We evaluate the suggested algorithm's performance using an event-driven simulator built in Python. Our algorithms are executed on a PC with an Intel 12700K processor and 32 GB of RAM. Two realistic network topologies, namely the Europe and German networks, are employed for the simulations, as depicted in Figure \ref{topologies}. Table \ref{tab:nru} lists these networks' physical characteristics. For both topologies under examination, $D_{avg}$ and $L_{avg}$ have distinct values. In both networks, we consider two bidirectional 4-core MCFs on each link. We assume that each core has a 4 THz capacity in the C band, which is further divided into 320 slots with a 12.5 GHz bandwidth each.  
Uncoupled-4 core MCF for long-haul propagation was recently developed and demonstrated by researchers \cite{takashi2024,uncoupled,trench,4core}. The trench-assisted refractive index profile is believed to cause very little inter-core crosstalk between neighboring cores. Thus, in contrast to crosstalk-aware and crosstalk-avoiding methods, we use the crosstalk-ignore spectrum assignment strategy, which results in higher spectrum utilization efficiency \cite{crosstalk}. However, the proposed algorithms are equally implementable to crosstalk-aware and crosstalk-avoid scenarios.

% \begin{figure}[htbp]
% \centering
% \includegraphics[width=0.9\linewidth]{4CF.png}
% \caption{Schematic diagram of the (a) cross-section of uncoupled 4-core MCF and (b) its trench-assisted refractive-index profile.}
% \label{fig:4CF}
% % \hspace{1mm}
% \rule{8.8cm}{0.1mm}
% \end{figure}

\begin{table}[htbp]
\centering
\caption{\bf Modulation Formats Against Path Length \cite{survey}}

\renewcommand{\arraystretch}{1.25} % Default value: 1

\begin{tabular}{cccc}
\hline
Modulation & $m$ & Supported Data   & Path Length \\
Format &  &  Rate ($\alpha$) in Gbps &($L_p$) in km \\
 \hline
DP-BPSK & 1 & 25 & 8000 \\
\hline
DP-QPSK & 2& 50 & 4000 \\
\hline
DP-8QAM & 3 & 75 & 2000 \\
\hline
DP-16QAM & 4 & 100 & 1000\\
\hline
DP-32QAM & 5 & 125 & 500 \\
\hline
DP-64QAM & 6 & 150 & 250 \\
\hline
\end{tabular}
  \label{tab:modulation}
\end{table}

Each node generates traffic with Poisson distribution with an average arrival rate of $\lambda$. An exponential distribution with a mean holding time of $1/\mu = T_h$ governs the holding time of the requests. As a result, the traffic load at the SDN controller equals $(\lambda/\mu) \times |V|$ Erlangs.
A request's required bandwidth is randomly selected from a uniformly distributed set $\delta = \{25, 50, 75, 100, 125, 150\}$. Additionally, the source and destination nodes are chosen randomly from the set $V$. Distance adaptive modulation levels are used, where Table \ref{tab:modulation} indicates the highest possible modulation level which can be used. A path is not considered if it is discovered to be longer than the maximum length allowed by the lowest modulation format, which is 8000 km in our case. To find alternative paths, the KSP-RMCSA, KDP-RMCSA and CALA-RMCSA algorithms use K = 3. The LB-RMCSA algorithm implements the shortest path whose weight is a combination of both link length and SOR and uses a single least-cost path for routing purposes.   
We insert an additional guard band of 12.5 GHz (1 spectrum slot) between two adjacent lightpaths. The first-fit core selection and first-fit spectrum assignment policy are used for core selection and spectrum assignment sub-problems in all the highlighted algorithms. Ten repetitions of each simulation are performed, and each simulation is run up to $10^5$ connection requests. The average results of ten iterations are presented with a 99$\%$ confidence interval. The first 10,000 requests are not recorded in order to obtain the results in steady-state.  

For comparison, the SP-RMCSA, KSP-RMCSA, KDP-RMCSA and LB-RMCSA \cite{weights} algorithms were implemented and used as benchmark algorithms. To expedite the route-finding process, all K candidate paths can either be (i) pre-computed and stored in memory or (ii) saved in run time after the first occurrence and used for future requests using the \emph{memoization} technique. In this work, we employ the latter approach. The LB-RMCSA algorithm could not take advantage of the \emph{memoization} technique to hasten the route-finding process since the link weights are periodically updated. In the LB-RMCSA algorithm, the weights parameter $\alpha$ is set equal to 0.5, and the weights are updated after serving 1500 connection requests to set a balance between blocking performance and service latency.

\subsection{Performance Metrics}
The RMCSA algorithms aim to effectively utilize the resources at hand and accept requests with diverse bandwidth demands in the presence of challenges imposed by constraints and bottleneck links. To assess the performance of the algorithms in terms of traffic admissibility, bandwidth admissibility, and resource utilization efficiency, we use three metrics: Request Blocking Probability (RBP), Bandwidth Blocking Probability (BBP), and Network Resource Utilization (NRU).
\begin{gather*}
             RBP= \frac{\textnormal{$R_b$}}{\textnormal{$R_b + R_a$}}, \\
             BBP= \frac{\textnormal{$\sum_{r\in R_b}b$}}{\textnormal{$\sum_{\forall r}b$}}, \\
            NRU = \frac{\textnormal{$\sum_{r\in R_a} SS_r \times H_p \times T_h$}}{\textnormal{$|E| \times |C| \times |S| \times$ $\tau$}}.
            % AHL = \frac{\textnormal{$\sum_{r\in R_a} H_p $}}{\textnormal{$R_a$}}.
\end{gather*}
The ratio of blocked requests to all requests that arrive is referred to as RBP. Meanwhile, BBP denotes the ratio of total bandwidth blocked over the total requested bandwidth. NRU indicates the proportion of the entire spectrum resources in the network used over a certain period of time. The list of symbols and notations used in this chapter is given in Table \ref{tab:symbols}. In this work, the total simulation time is taken as the total observation time $\tau$.
 
In future networks \cite{latency}, reducing delay is heavily emphasized to support real-time applications such as eXtended Reality (XR). The time it takes for the RMCSA algorithm to set up a lightpath (generate $P_k, C, SSI, SS_r$ and $m$) following a request at the SDN controller is measured as the resource finding time. The Average Service Latency (ASL) metric has been employed to measure the delay in relation to the latency in the resource-finding process. 
\begin{gather*}
             ASL = \frac{\textnormal{$\sum_{r\in R_a} (\Delta T_s) $}}{\textnormal{$R_a$}},\\
             \Delta T_s = T_{allocated} - T_{arrived}.
\end{gather*}
$T_{arrived}$ indicates the time when the request arrives at the SDN controller, whereas $T_{allocated}$ indicates when the resources needed to set up the lightpath have been found by the RMCSA algorithm. 

\subsection{Results and Discussion} 

\begin{figure}[ht]
\captionsetup[subfigure]{justification=centering}
    \centering
      \begin{subfigure}{0.4\textwidth}
        \includegraphics[width=\textwidth]{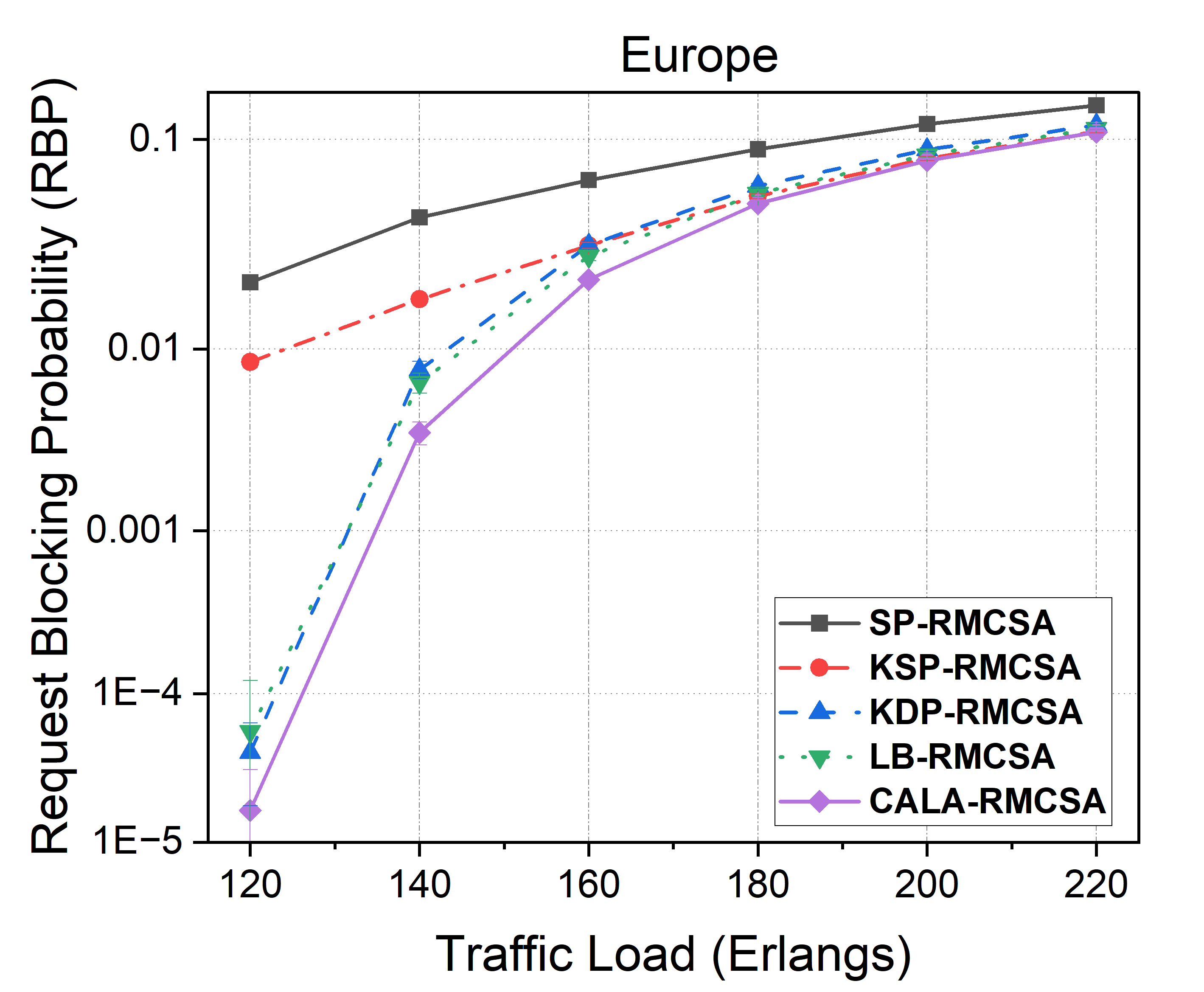}
          \caption{}
          \label{fig:RBP_E}
      \end{subfigure}
      \begin{subfigure}{0.4\textwidth}
        \includegraphics[width=\textwidth]{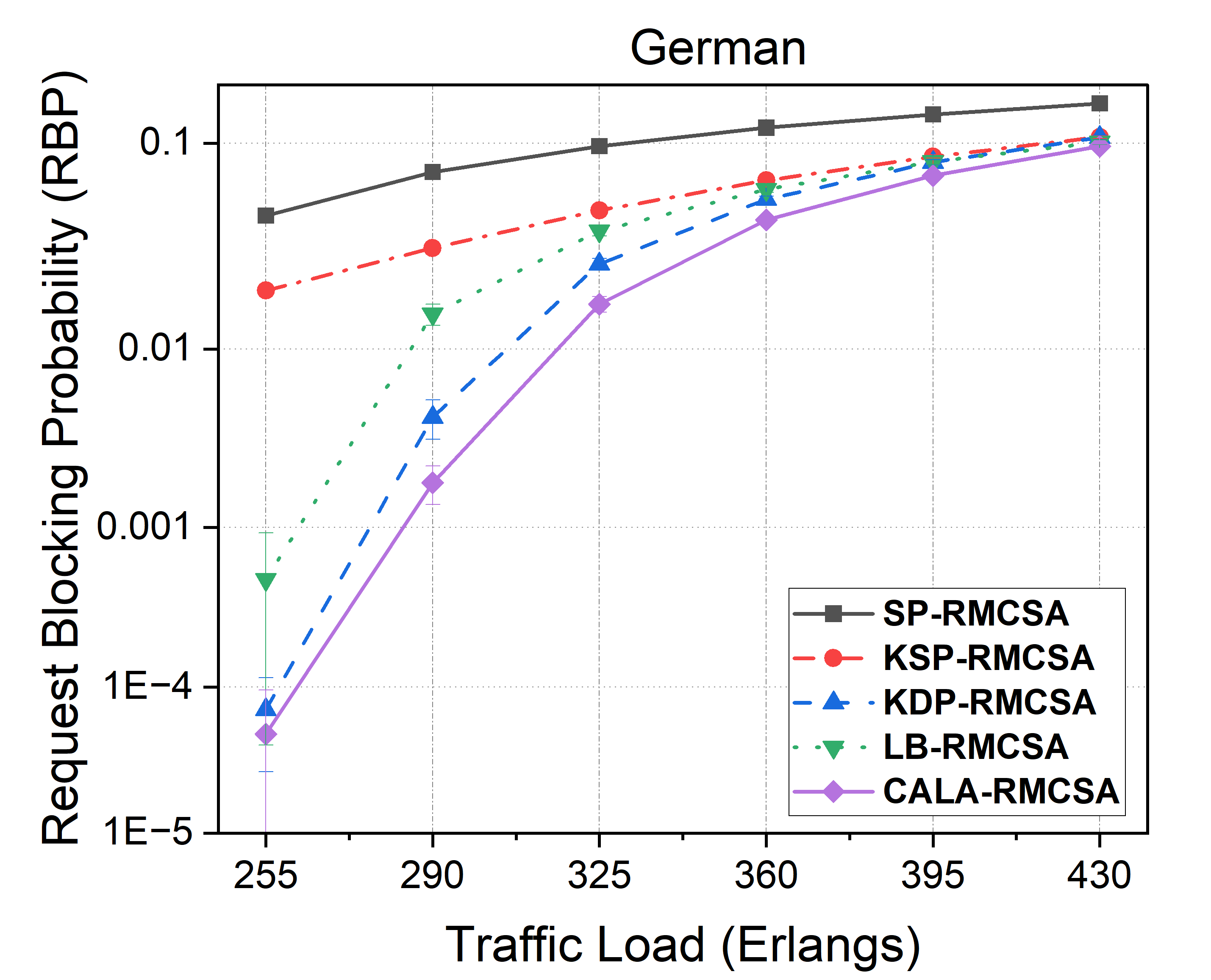}
          \caption{}
          \label{fig:RBP_G}
      \end{subfigure}
      % \begin{subfigure}{0.49\textwidth}
      %   \includegraphics[width=\textwidth]{RBP_U.png}
      %     \caption{}
      %     \label{fig:RBP_U}
      % \end{subfigure}
    \caption{Request blocking probability of (a) Europe network, and (b) German network.}
\label{RBP}
% \rule{18.3cm}{0.1mm}
\end{figure} 

The plots in Figure \ref{RBP} present the RBP at various traffic load values derived from simulation outcomes across Europe and German network topologies. As anticipated, the SP-RMCSA algorithm demonstrates the least favorable RBP performance among the algorithms under examination. This outcome can be attributed to the routing algorithm's sole focus on assessing the availability of required slots only along the shortest path, without displaying any awareness of traffic conditions. On the other hand, the KSP-RMCSA algorithm exhibits enhanced RBP performance when compared to the SP-RMCSA algorithm by scrutinizing resource availability across three alternative shortest paths.

It is worth noting that in the German network, the KDP-RMCSA algorithm capitalizes on a higher average nodal degree ($D_{avg}$ = 3.05 $>$, where $D_{avg}$ $>$ K), resulting in superior performance compared to the KSP-RMCSA algorithm with an equal number of alternative candidate paths. Conversely, in the Europe network with $D_{avg}$ = 2.93, where $D_{avg}$ $<$ K, the KDP-RMCSA algorithm outperforms the KSP-RMCSA algorithm solely for low traffic load values. However, under high loads (180, 200, $\&$ 200 Erlangs), the KDP-RMCSA algorithm yields higher RBP in comparison to the KSP-RMCSA algorithm. These findings highlight the dependence of the KDP-RMCSA algorithm's performance on $D_{avg}$.
 
The LB-RMCSA algorithm computes the shortest path by considering new link weights ($NLW_l$), which take into account both the lengths of the individual links and their spectrum occupancy ratios. In addition to surpassing the performance of the SP-RMCSA algorithm, the LB-RMCSA algorithm also outperforms the KSP-RMCSA algorithm, which utilizes three alternative paths. Furthermore, in contrast to the KDP-RMCSA algorithm, which results in increased blocking under high traffic loads in the Europe network, the LB-RMCSA algorithm excels over the benchmark SP-RMCSA and KSP-RMCSA algorithms at both lower and higher traffic values for both test networks. Hence, by incorporating spectrum occupancy information of the network's links, the LB-RMCSA algorithm can identify the shortest routes with high acceptance probabilities.

The novel CALA-RMCSA algorithm consistently outperforms SP-RMCSA, KSP-RMCSA, KDP-RMCSA, and LB-RMCSA in the context of Europe and German network topologies, regardless of whether the traffic load is low or high. Specifically, in the case of the Europe network, the CALA-RMCSA algorithm demonstrates an 80.6$\%$, 14.9$\%$, 16.0$\%$, and 9.3$\%$ reduction in average RBP when compared to the SP-RMCSA, KSP-RMCSA, KDP-RMCSA, and LB-RMCSA algorithms, respectively. Meanwhile, when implemented on the German network topology, the CALA-RMCSA algorithm achieves substantial RBP reductions, 62.6$\%$ lower than the SP-RMCSA algorithm, 36.2$\%$ lower than the KSP-RMCSA algorithm, 15.8$\%$ lower than the KDP-RMCSA algorithm, and 22.9$\%$ lower than the LB-RMCSA algorithm. The simulation results unequivocally establish the superior performance of our proposed CALA-RMCSA algorithm over all benchmark algorithms.

\begin{figure}[ht]
\captionsetup[subfigure]{justification=centering}
    \centering
      \begin{subfigure}{0.4\textwidth}
        \includegraphics[width=\textwidth]{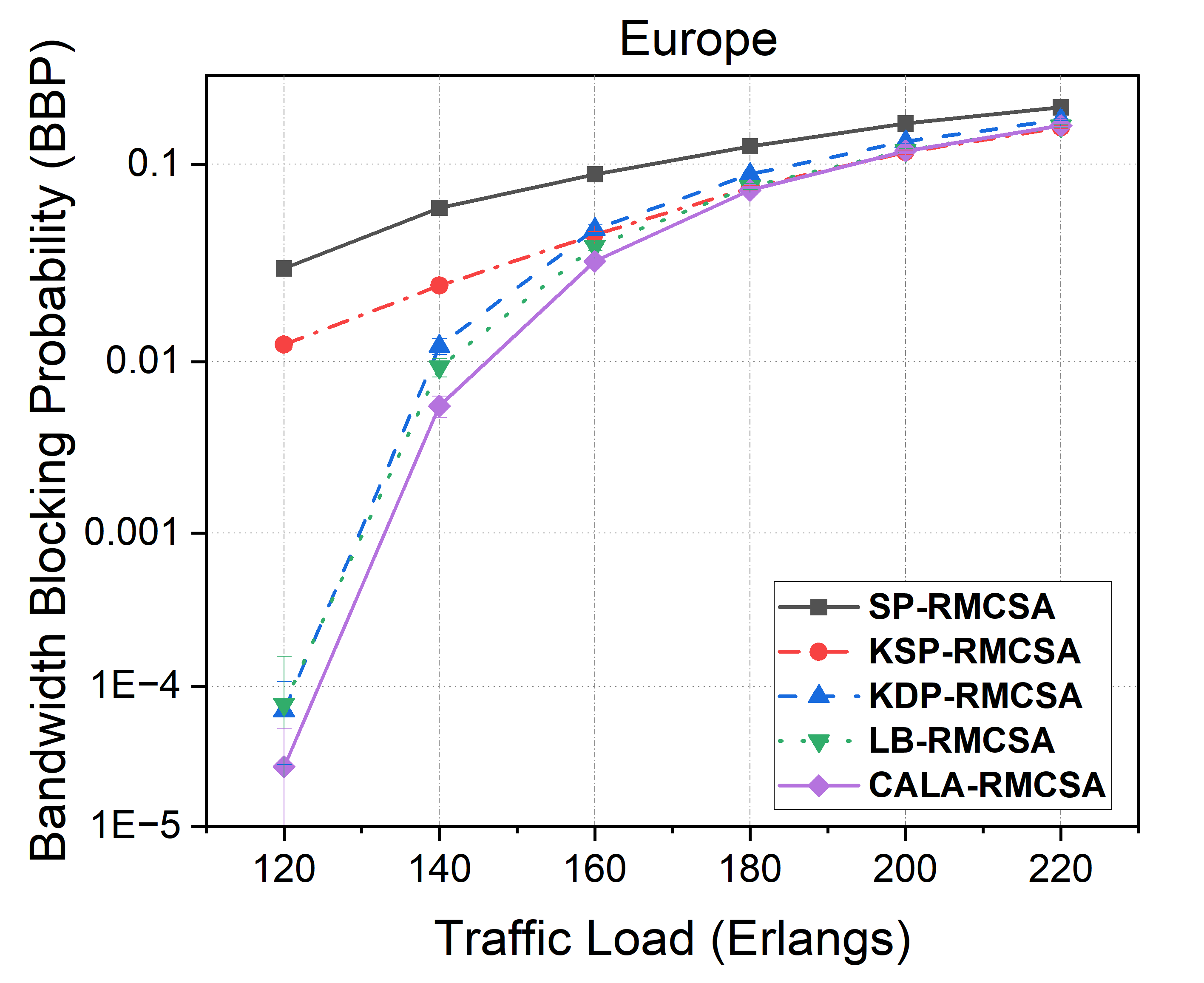}
          \caption{}
          \label{fig:BBP_E}
      \end{subfigure}
      \begin{subfigure}{0.4\textwidth}
        \includegraphics[width=\textwidth]{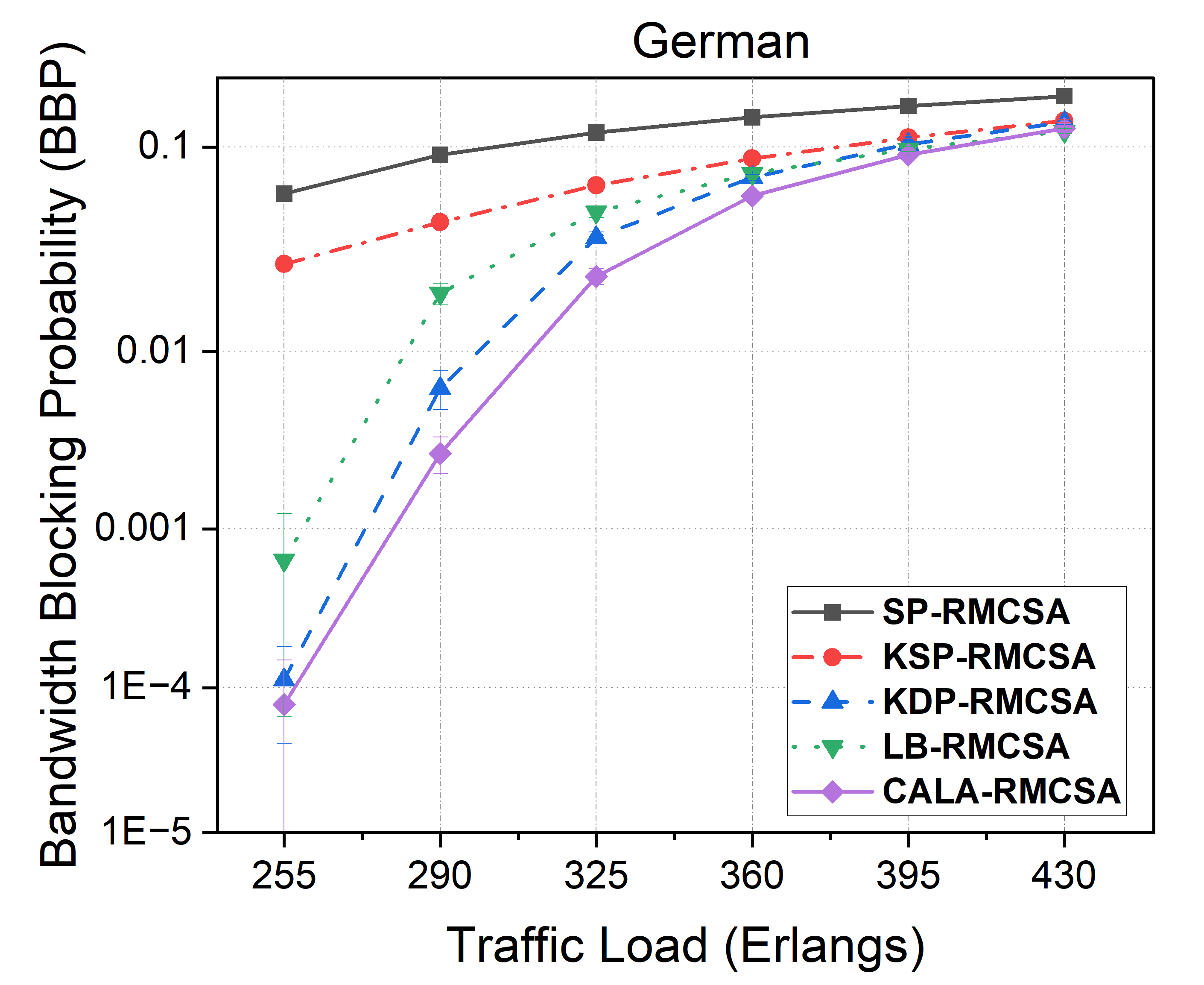}
          \caption{}
          \label{fig:BBP_G}
      \end{subfigure}
      % \begin{subfigure}{0.49\textwidth}
      %   \includegraphics[width=\textwidth]{RBP_U.png}
      %     \caption{}
      %     \label{fig:RBP_U}
      % \end{subfigure}
    \caption{Bandwidth blocking probability of (a) Europe network, and (b) German network.}
\label{BBP}
% \rule{18.3cm}{0.1mm}
\end{figure}

Since blocking of high bit-rate requests has a higher impact on the network's data transmission capacity, measuring the BBP for multi-rate connection requests is important. Figure \ref{BBP} illustrates the BBP performance of the algorithms under consideration for both networks examined. All evaluated algorithms demonstrate comparable performance to the RBP case, with the proposed CALA-RMCSA algorithms consistently achieving the lowest BBP values across all traffic load scenarios. These findings indicate that the proposed algorithm enhances RBP by effectively utilizing traffic information without compromising BBP.

  \begin{figure}[ht]
\captionsetup[subfigure]{justification=centering}
    \centering
      \begin{subfigure}{0.4\textwidth}
        \includegraphics[width=\textwidth]{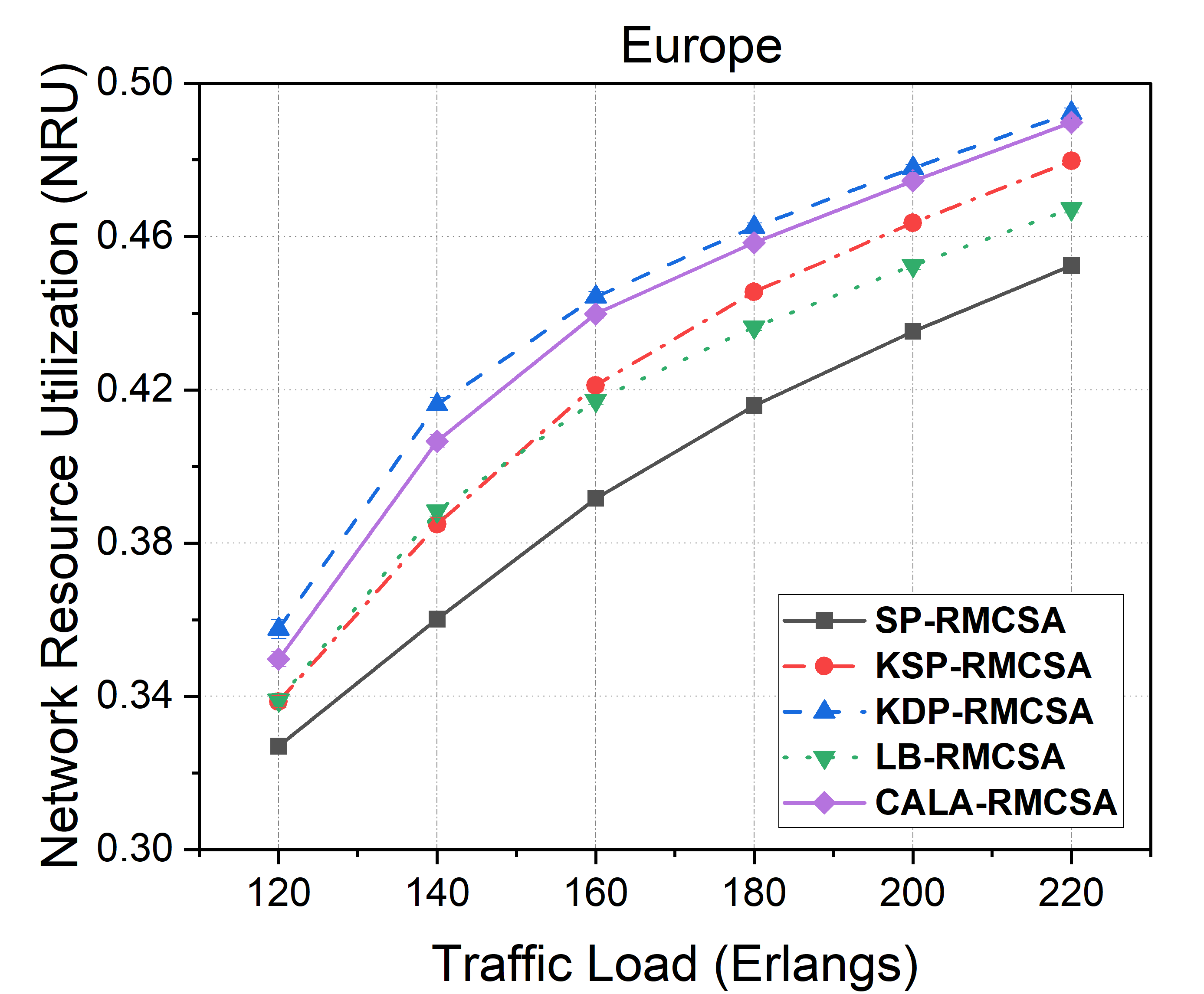}\hfill
          \caption{}
          \label{fig:NRU_E}
      \end{subfigure}
      \begin{subfigure}{0.4\textwidth}
        \includegraphics[width=\textwidth]{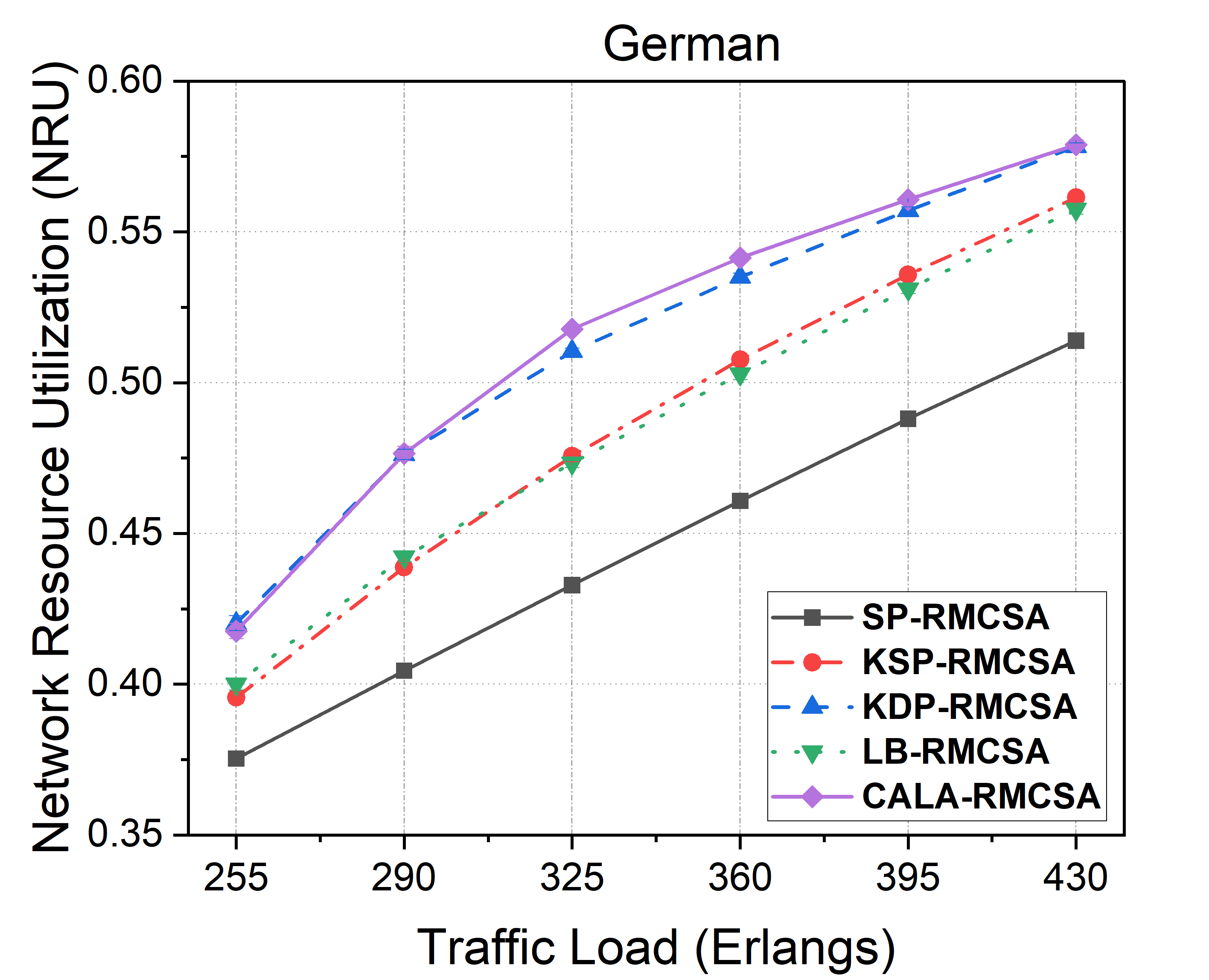}\hfill
          \caption{}
          \label{fig:NRU_G}
      \end{subfigure}
      % \begin{subfigure}{0.49\textwidth}
      %   \includegraphics[width=\textwidth]{NRU_U.png}
      %     \caption{}
      %     \label{fig:NRU_U}
      % \end{subfigure}
    \caption{Network resource utilization of (a) Europe network, and (b) German network topologies.}
\label{NRU}
% \rule{18.3cm}{0.1mm}
\end{figure}

The performance of the proposed RMCSA algorithm and the benchmark RMCSA algorithms in terms of Network Resource Utilization (NRU) is depicted in Figure \ref{NRU}. Across German networks, the proposed CALA-RMCSA algorithm consistently exhibits higher NRU values than all the benchmark algorithms. For the Europe network topology, the CALA-RMCSA algorithm demonstrates higher NRU at all load values when compared to the SP-RMCSA, KSP-RMCSA, and LB-RMCSA algorithms. These observations indicate that the proposed CALA-RMCSA algorithm can capitalize on underutilized spectral resources, a capacity not fully realized by benchmark algorithms due to congestion issues associated with highly central links.

\begin{figure}[ht]
\captionsetup[subfigure]{justification=centering}
    \centering
      \begin{subfigure}{0.4\textwidth}
        \includegraphics[width=\textwidth]{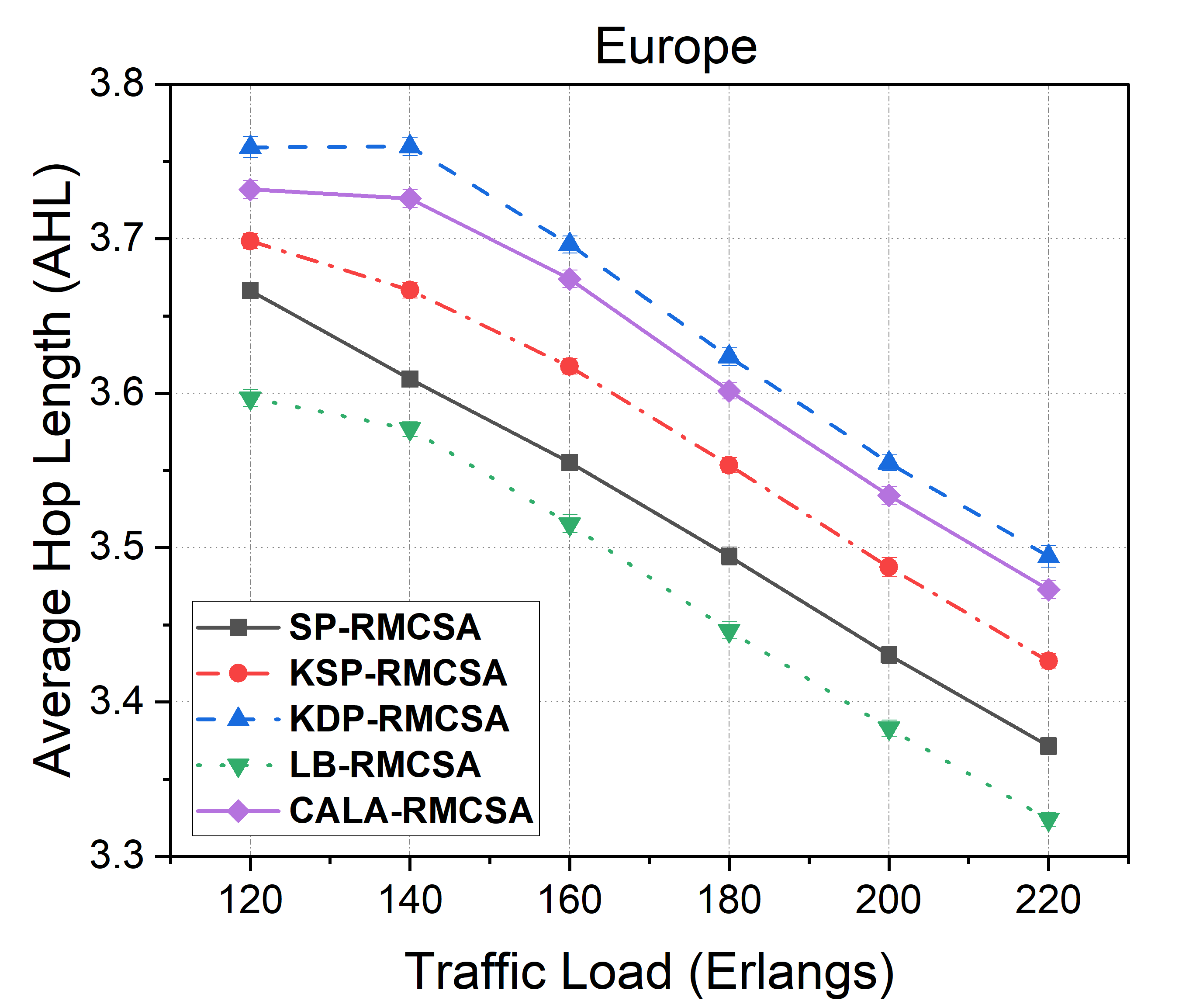}\hfill
          \caption{}
          \label{fig:AHL_E}
      \end{subfigure}
      \begin{subfigure}{0.4\textwidth}
        \includegraphics[width=\textwidth]{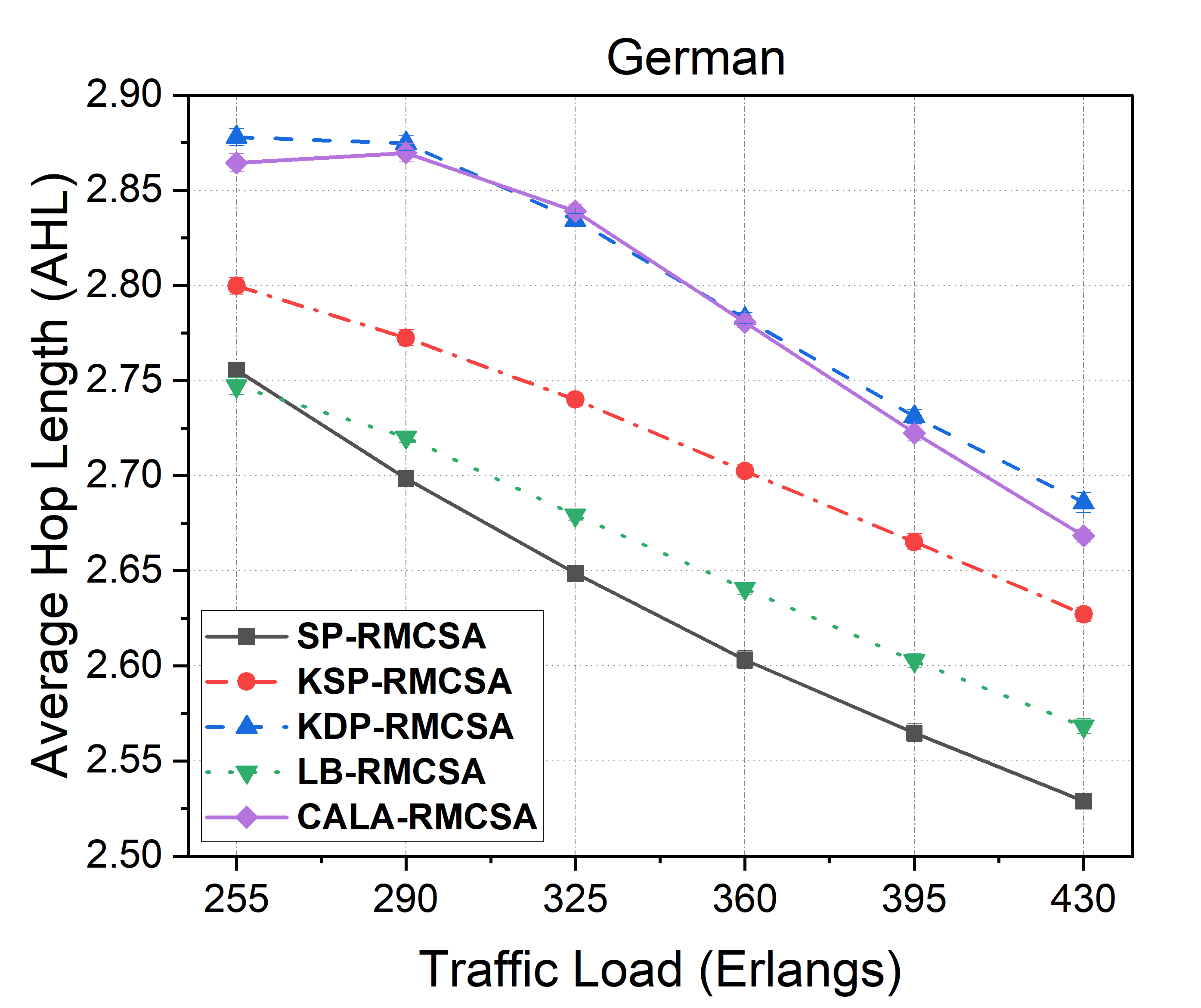}\hfill
          \caption{}
          \label{fig:AHL_G}
      \end{subfigure}
      % \begin{subfigure}{0.49\textwidth}
      %   \includegraphics[width=\textwidth]{APL_U.png}
      %     \caption{}
      %     \label{fig:APL_U}
      % \end{subfigure}
    \caption{Average hop length of (a) Europe network, and (b) German network topologies.}
\label{fig:AHL}
% \rule{18.3cm}{0.1mm}
\end{figure}

Notably, the KDP-RMCSA algorithm yields the highest NRU values in the case of the Europe network. This is because the KDP-RMCSA algorithm tends to select longer paths that span a greater number of links, while the CALA-RMCSA algorithm opts for shorter alternative paths, which involve fewer links. To provide empirical support for this observation, we have calculated the average number of hops in all the accepted requests for all algorithms using the formula:
$$AHL = \frac{\sum_{r\in R_a} H_p}{R_a}$$
Average Hop Length (AHL) denotes the average number of links/hops in all accepted requests. The values of $AHL$ for all the algorithms under consideration are shown in Figure \ref{fig:AHL}. The results presented in Figure \ref{fig:AHL} clearly illustrate that the KDP-RMCSA algorithms tend to select paths with a greater number of hops in comparison to the CALA-RMCSA algorithm. The value of NRU of an accepted request is directly proportional to the number of hops in its working path. Consequently, the established lightpaths occupy more spectrum slots, as indicated by the elevated NRU values in Figure \ref{NRU}. 

Furthermore, it is important to note that the higher NRU achieved by the KDP-RMCSA algorithm compared to the proposed CALA-RMCSA algorithm does not result in a commensurate reduction in RBP, as demonstrated in Figure \ref{RBP}. In light of these findings, it can be concluded that the proposed CALA-RMCSA algorithm is a more favorable choice, as it not only leads to a lower RBP when compared to the KDP-RMCSA algorithm but also conserves a greater amount of spectrum resources for accommodating future connection requests.

  \begin{figure}[ht]
\captionsetup[subfigure]{justification=centering}
    \centering
      \begin{subfigure}{0.4\textwidth}
        \includegraphics[width=\textwidth]{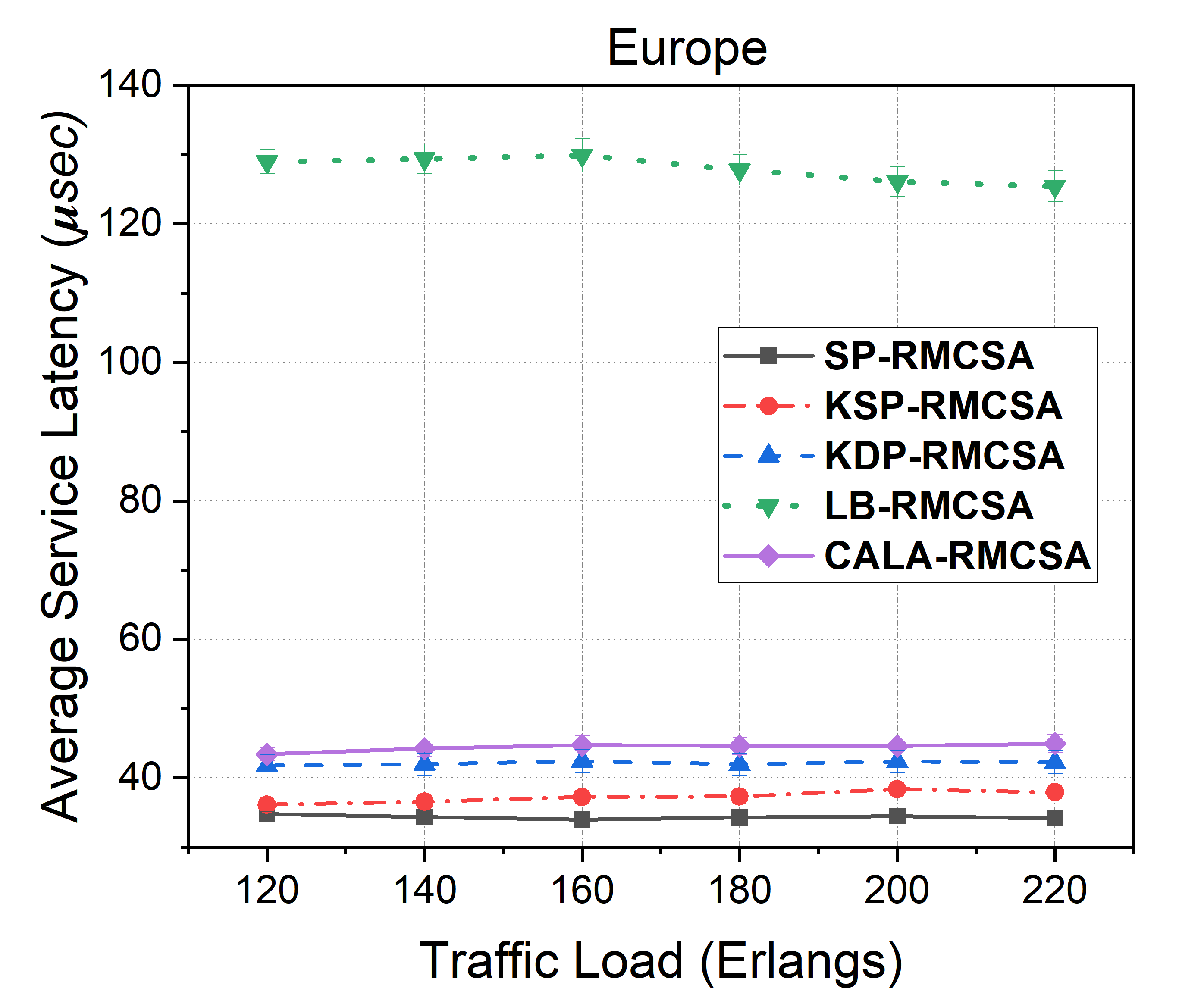}\hfill
          \caption{}
          \label{fig:del_E}
      \end{subfigure}
      \begin{subfigure}{0.4\textwidth}
        \includegraphics[width=\textwidth]{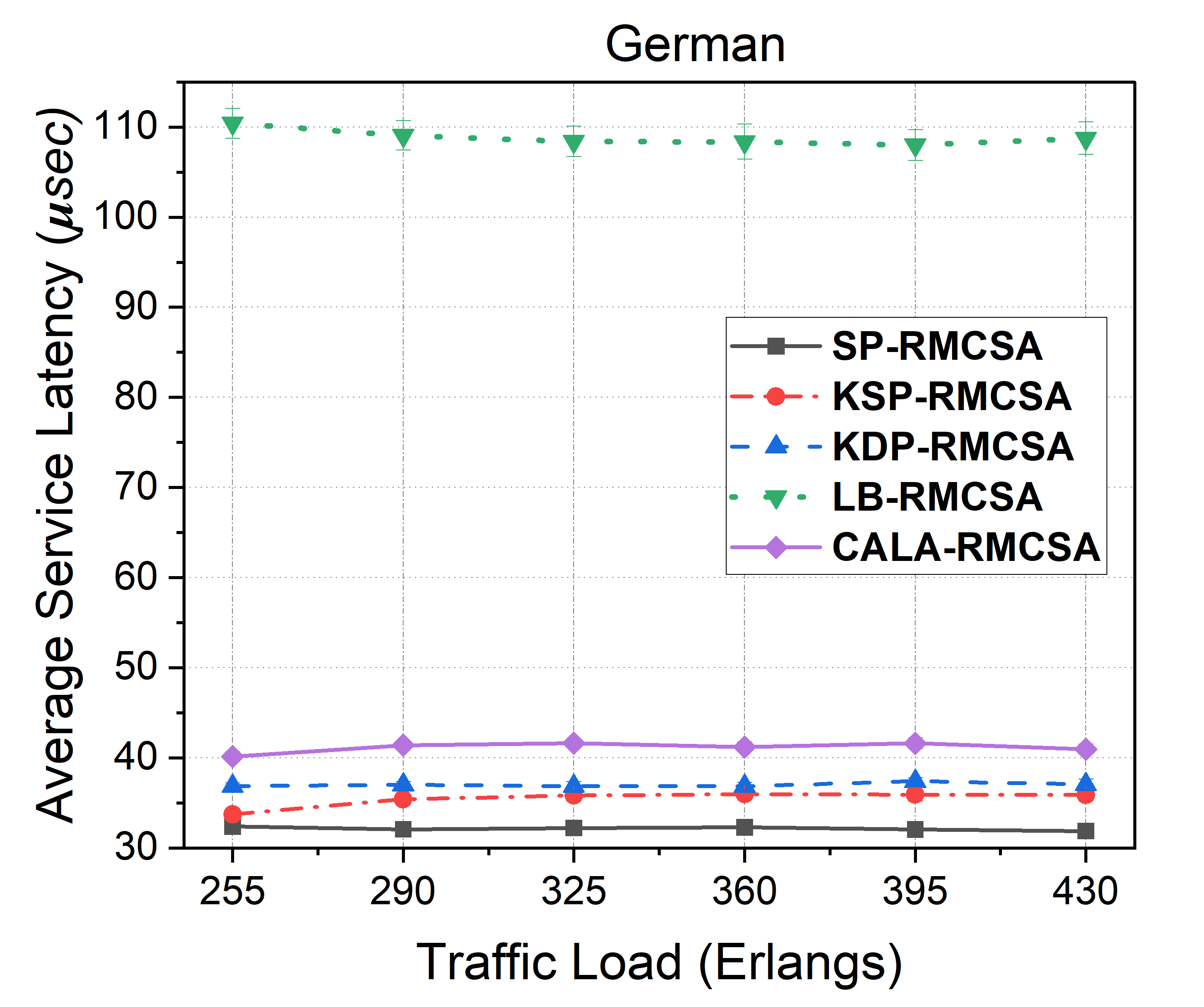}\hfill
          \caption{}
          \label{fig:del_G}
      \end{subfigure}
      % \begin{subfigure}{0.49\textwidth}
      %   \includegraphics[width=\textwidth]{ASL_U.png}
      %     \caption{}
      %     \label{fig:del_U}
      % \end{subfigure}
    \caption{Average service latency in $\mu sec$ for (a) Europe network, and (b) German network topologies.}
\label{fig:ASL}
% \rule{18.3cm}{0.1mm}
\end{figure}

Figure \ref{fig:ASL} shows the average service latency in each algorithm's search for the required resources. The SP-RMCSA, KSP-RMCSA, and KDP-RMCSA algorithms exhibit relatively lower $ASL$, as demonstrated by simulation results. This is because all the shortest candidate paths between each $s$-$d$ pair remain the same. Once computed, they are reused for future connection requests using the implemented novel cache. As expected, The KSP-RMCSA and KDP-RMCSA algorithms exhibit slightly higher ASL when compared to the SP-RMCSA algorithm. This is due to the fact that both KSP-RMCSA and KDP-RMCSA algorithms assess resources on three shortest candidate paths, while the SP-RMCSA algorithm examines the same only on the first shortest path.

The LB-RMCSA algorithm demonstrates significantly high $ASL$ values out of all the test networks. Candidate routes between the same $s$-$d$ pairs must be recomputed because the pre-computed and stored paths must be erased from memory in order to perform the periodic update of link weights. The LB-RMCSA algorithm exhibits 188$\%$ and 164$\%$ higher $ASL$ compared to the proposed CALA-RMCSA algorithm for Europe and German topologies, respectively. Therefore, the LB-RMCSA algorithm may not be suitable for $5^{th}$ Generation optical networks with stringent service latency requirements. 

A comparison of the KSP-RMCSA and KDP-RMCSA algorithms in the proposed CALA-RMCSA algorithm reveals a marginally higher $ASL$, attributed to the dynamic route-finding process inherent in the proposed CALA-RMCSA algorithm. The CALA-RMCSA algorithm uses the advantage of having pre-computed routes saved in memory for later use. It fetches all pre-computed paths from memory because the link weights are fixed, as described in section \ref{caching}. 

From the simulation results discussed above, we can say that the proposed resource allocation algorithm performs better than the benchmark algorithms regarding request blocking probability as well as resource utilization efficiency. Fortunately, it does not show any substantial limitations in terms of delay in the request establishments. In summary, the suggested CALA-RMCSA algorithm shows a promising combination of performance among all performance metrics. Therefore, the CALA-RMCSA represents a significant leap forward in addressing the stringent capacity, flexibility, and latency requirements of routing and spectrum allocation in future SDM-EONs.

\section{Conclusion}
In this work, we demonstrate a novel RMCSA algorithm for dynamic resource allocation in SDM-EONs. The algorithm assesses the current traffic conditions to make informed decisions on finding alternative routes with a higher probability of acceptance. The proposed CALA-RMCSA algorithm employs a mechanism to circumvent the presence of bottleneck links on a given route by identifying and utilizing alternative routes that do not share congested links. Further, to reduce the service delay time, the proposed heuristic algorithm incorporates a sophisticated caching system to expedite the process of determining alternate routes to minimize the average delay in accepting connection requests. Extensive simulations were conducted on two realistic network typologies for dynamic traffic. The algorithm under consideration is evaluated on SDM-EONs, which can allocate bandwidth flexibly to accommodate heterogeneous bandwidth requirements. The simulation results demonstrate that our proposed CALA-RMCSA algorithm outperforms the benchmark SP-RMCSA, KSP-RMCSA, KDP-RMCSA, and LB-RMCSA algorithms to reduce connection blocking and improve spectrum utilization efficiency.
Moreover, the method under consideration eliminates the necessity of regular updates to the link weights for load distribution across network links. The suggested methodology effectively utilizes a cache to store previously computed paths, reducing overall time and computational complexity. In contrast, the LB-RMCSA algorithm, at its current level of advancement, necessitates periodic updates of link weights in order to mitigate congestion-induced blocking. Therefore, the LB-RMCSA algorithm effectively mitigates blocking by imposing a noticeably higher delay in accepting connection demands than the proposed CALA-RMCSA algorithm.
The suggested CALA-RMCSA enhances the performance by integrating link congestion and latency awareness into its core design. The dynamic routing algorithm proposed herein also exhibits resilience by employing a single link-disjoint path. As a result, it enables more efficient spectrum utilization, enhances reliability, and supports communication services sensitive to service latency. Consequently, it becomes a valuable asset for network operators in the era of data-driven telecommunications, complementing the existing range of application-specific resource allocation algorithms. 

\subsection*{Disclosures} The authors declare no conflicts of interest.

\end{document}